\documentclass[sn-basic]{sn-jnl}% Basic Springer Nature Reference Style/Chemistry Reference Style

\usepackage{graphicx}%
\usepackage{multirow}%
\usepackage{amsmath,amssymb,amsfonts}%
\usepackage{amsthm}%
\usepackage{mathrsfs}%
\usepackage[title]{appendix}%
\usepackage{xcolor}%
\usepackage{textcomp}%
\usepackage{manyfoot}%
\usepackage{booktabs}%
\usepackage{algorithm}%
\usepackage{algorithmicx}%
\usepackage{algpseudocode}%
\usepackage{listings}%
\usepackage{lscape}
\usepackage{footnote}
\usepackage{tablefootnote}
\usepackage[english]{babel}

\usepackage{xspace}
\newcommand{\kms}{\ensuremath{\,\mathrm{km\, s^{-1}}}\xspace}

\begin{document}

\title[Explosive Molecular Outflows]{Explosive Molecular Outflows}
%:Remnants of the Violent Formation of the Massive Stars}


\author*[1]{\fnm{Luis A.} \sur{Zapata}}\email{l.zapata@irya.unam.mx}



\affil*[1]{\orgdiv{Instituto de Radioastronomía y Astrofísica}, \orgname{Universidad Nacional Autónoma de México}, \orgaddress{\street{Campus Morelia}, \city{Morelia}, \postcode{58089}, \state{Michoacán}, \country{México}}}

%\affil[2]{\orgdiv{Department}, \orgname{Organization}, \orgaddress{\street{Street}, \city{City}, \postcode{10587}, \state{State}, \country{Country}}}

%\affil[3]{\orgdiv{Department}, \orgname{Organization}, \orgaddress{\street{Street}, \city{City}, \postcode{610101}, \state{State}, \country{Country}}}


\abstract{About fifteen years ago a new type of extreme (with kinetic energies of $\mathrm{E}_k \sim 10^{47-49}\, \mathrm{erg}$) molecular outflows associated with very luminous ($\geq 10^5\, \mathrm{L}_\odot$) and massive (10$^3$ M$_\odot$) young clusters was confirmed, the \textit{Explosive Molecular Outflows}. This new class of outflows is largely different from the 
classical bipolar protostellar flows, with spatial distributions made of numerous narrow straight filament- or streamer-like ejections in an almost isotropic arrangement and with clear Hubble--Lemaître-like expansion motions.  Straight filaments point directly to the center of expanding molecular or ionized shells, which exhibit expansion velocities of about 10--50\kms. However, no young massive stars are clearly located there, probably because they moved to other places. 
These physical characteristics suggest that explosive outflows are short-lived in nature and possibly generated by an energetic single and brief disrupting event.
The most up-to-date theoretical model for explaining their nature involves the disruption of non-hierarchical massive protostellar systems, where members may either form a close binary (with separations of a few au) or merge into a single massive star, as recently proposed for the nearest high-mass star-forming region, Orion BN/KL. The present-day number of discovered explosive outflows and their estimated duration of some thousand years have set this rate (one every 140 years) to a value similar to the Galactic supernova rate and the Galactic rate of formation of massive stars, suggesting that this mode of formation likely is occurring in a large number of the massive stars. Future sensitive and detailed observations (e.g.\ with the James Webb Space Telescope, Atacama Large Millimeter/Submillimeter Array or in a near future the next generation Very Large Array) to a large number of close-by high-mass star forming regions could reveal many more of these remarkable flows and their true nature.}

\keywords{ISM: kinematics and dynamics, Stars: winds, outflows, Stars: formation}

\maketitle

\setcounter{tocdepth}{3} % TOC subsubsections
\tableofcontents

\newpage
\section{Formation of the massive stars}\label{sec1}

\subsection{Theoretical and observational challenges}\label{subsec1}

One of the unresolved problems in modern astrophysics is the formation of the massive stars, that is, stars with masses larger than 8 M$_\odot$, where a heavy Fe core is surely formed near the end of the star's life. These massive stars have luminosities thousand of times the luminosity of the Sun, for example, a 60 M$_\odot$ main-sequence star will have a luminosity of 8.0 $\times$ 10$^4$ L$_\odot$ and will have a very short lifetime of only a few million years \citep{massey2013}. Massive stars are rare indeed, with only 2 to 3 forming for every 1,000 low- or intermediate-mass \citep{2025eks}. The death of these giant stars produces powerful supernovae (with kinetic energies reaching the 10$^{52}$ erg) generating a large fraction of the heavy elements and injecting a large amount of turbulent energy into the interstellar medium, which may trigger the formation of new stars \citep{gor2004}. However, at this moment, there is evidence that some massive stars will probably collapse into black holes without a supernova remnant \citep{rey2015}. Massive stars exhibit a higher multiplicity fraction compared to their lower-mass counterparts. For instance, \citet{Lan2023} found a multiplicity fraction of 0.6 in an optical survey of more than 100 O-type stars, meaning that 60\% of the observed targets were confirmed to be part of a multiple stellar system. This survey is in agreement with other surveys made for example in the Southern Hemisphere \citep{sana2014}, where the fraction of massive stars with at least one companion within two hundred milli-arcsec (at a typical distance of 2 kpc this corresponds 400 au) is about 0.53. In addition, these surveys reported very close binaries (eclipsing binaries) or multiple systems with separations of only 0.5 milli-arcsecs (at typical distances of 2 kpc). As discussed in \cite{Lan2023}, studying the multiplicity of massive stars should provide value information for the physical formation models.   

The first proposed problem with the formation of the massive stars appeared about 50 years ago.  \cite{Lar1971,Kan1974, york1977} found that a forming star that reaches a mass of 8 M$_\odot$ will have enough luminosity that will stop the spherical mass accretion onto the star since the Eddington limit --calculated based on dust opacity rather than the much smaller free-free opacity-- was reached. This implies that stars with masses superior to 8 M$_\odot$ will not form by spherical mass accretion. However, later works as those of \cite{wolf1987,nak1989} proposed that stars with larger masses will be formed if the mass accretion onto the protostars is not spherical. If flattened accreting disks are formed because of the inherit rotation from their parents molecular clouds, the part of radiation pressure impacting the dust grains can escape out from the poles forming powerful bipolar outflows. More recent works as \cite{ben2001,yorke2002} improving the radiation transfer models and using hydrodynamic codes found that stars with masses up to 42.9 M$_\odot$ could be formed via circumstellar disks. Moreover, multi-dimensional radiation hydrodynamic simulations had yielded the growth of the highest-mass stars ever formed with final masses of 137.2 M$_\odot$ \citep{kuiper2010}. In conclusion, many of the up-to-date radiation-hydrodynamic simulations of high-mass star formation indicate that radiation pressure does not represent an important trouble for the formation of the massive stars with surrounding gas and dusty disks up to several hundreds of solar masses \citep{krum2015}.  

The second suggested problem is the thermal pressure of the ionized gas. Early OB spectral type stars produce a large amount of Lyman continuum radiation 
that is able to ionize the surrounding molecular gas forming an H\,II region around the massive star. The molecular mass accretion onto the massive protostar should then be reversed since the thermal pressure of the ionized gas is higher (with a temperature of 10$^4$ K) than the thermal pressure of the molecular gas (with a temperature of 100 K). The pressure therefore is dominated by the ionized gas that should expand quickly. The solution to this problem came from the ionization budget in the vicinities of the massive star. Higher accretion rates imply high mass densities in the surroundings of the massive star, which significantly reduces the Str\"omgren radius. If the Str\"omgren radius is very small so that the escape speed from its outer edge is smaller than 10\kms (the sound speed of the ionized gas at a temperature of 10$^4$ K), the ionized gas irreversibly will be infalling directly to the young stars \citep{wal1995,keto2002,keto2003,keto2006,thomas2010,tan2014, beu2007}.  
Finally, the third problem with the formation of the massive stars is the short zero main sequence time-scale of the massive stars compared to the accretion time-scales \citep{nak1989,ben2001}. As mentioned before, a massive star with a mass of 60 M$_\odot$ (or O spectral-type star) has a zero main sequence time-scale of only a few 100,000 of years, this implies that the accretion time-scale should be very short with high accretion rates. To form a massive star with 60 M$_\odot$, it will need an accretion rate of 6$\times$10$^{-4}$ M$_\odot$ years$^{-1}$ in 10$^5$ years. This has been proposed by \cite{mckee2003}, a massive star of 20 M$_\odot$ is formed in 10$^5$ years with accretion rates between 10$^{-4}$ to 10$^{-3}$ M$_\odot$ years$^{-1}$. 
Such large accretion rates have already been reported in massive star forming regions, G24.78+0.08 \citep{bel2006}, G31.41+0.31 \citep{2018beltran},  W51 North \citep{zapata2008}, NGC 7538S \citep{san2010,mel2012}, G19.61-0.23 \citep{furuya2011}, JCMT 18354-0649S \citep{liu2011}, IRAS 18360$-$0537 \citep{qiu2012},  NGC 7538 IRS1 \citep{beuther2013,2020san}, and four other massive star forming regions \citep{yue2021}. To these short dynamical timescales, one must add the time spent in massive infrared dark clouds during low- to intermediate-mass stages \citep{san2019,svo2019,san2017,wang2014,zhang2011,zhang2009}. 

Along with the previously mentioned issues for the formation of massive stars, we need to add
the observational difficulties. As massive stars only comprised a very small amount ($\sim$ 1\%) of the total stellar mass spectrum \citep{kooupa2001}, mapping their evolutionary path from protostellar massive cores to the main sequence remains challenging \citep{2018motte}. The first of these observational troubles is the large distances of the regions where massive stars are forming, for an example, the closest massive star forming region is the Orion Molecular Cloud (OMC) located to a distance of 417$\pm$7 parsecs \citep{men2007}, and one of the farthest regions is Westerhout 49 (also known as W49), located at 
a distance of 11.1$^{+0.79}_{-0.69}$ parsecs \citep{zhang2013}. Therefore, this becomes an observational challenge. If one wishes to reveal a dusty circumstellar disk surrounding a massive protostar with a size of 500 au at a distance of 5 kpc, an angular resolution better than 0.1$''$ is needed. Observations with radio and millimeter/submillimeter interferometers are one of the best options. Optical and infrared telescopes also offer similar angular resolutions; however, the large extinctions A$_v$ = 10$-$1000 at these wavelengths do not allow to trace the embedded circumstellar disks, particularly in the optical regime. Star forming regions with values of mass surface densities of $\Sigma$ $\sim$ 1 gr cm$^2$ present these large extinctions \citep[][]{KauffmannPillai2010,ButlerTan2012}. 

Observational searches have been conducted to find evidence of circumstellar disks around massive protostars. At this moment there is more evidence of disks surrounding 
high mass B-type stars as for example in IRAS 20126$+$4140 \citep{cesa1999,zhang1998,cesa2005,tk2005,jin2012,chen2016}, Cepheus A HW2 \citep{nimesh2005,jimenez2007} AFGL 490 \citep{sch2002,sch2006}, G192.16$-$3.82 \citep{shep2001,shep2004}, HH 80-81-mm \citep{manuel2011,carlos2012,girart2017,2020anez,manuel2023}. These disks have dimensions that go from a few hundred to thousands of au and masses that range 
from a few to tens of solar masses. Dedicated searches have also been carried out on larger samples of OB young stars. \cite{zap2006} using Very Large Array (VLA) 7 mm continuum  observations toward a sample of 10 luminous Infrared Astronomical Satellite (IRAS) sources found that
the millimeter and centimeter emission is either tracing H\,II regions, thermal jets or circumstellar disks/envelopes. Similar results were reported by \cite{gib2004}. Approximately of even more massive protostars, young O-type stars, were first reported to have been associated with very large ($\sim$10$^4$ au) rotating toroids with masses of more than 10 M$_\odot$ \citep{cesaroni2007,beltran2016}.  These toroids are likely forming a cluster of stars in their centers. Observations with a higher angular resolution have allowed to reveal smaller Keplerian circumstellar disks surrounding O-type protostars with sizes of a few hundred of au, see for example G35.20$-$0.74 N \citep{2013alvaro}, NGC 7538 IRS1 \citep{2015goddi}, IRAS 16547$-$4247 \citep{zapata2015,zapata2019,zap2024,tanaka2020}, G11.92$-$0.61 MM1 \citep{2016ilee,2018ilee}, NGC 6334 I(N) SMA1b \citep{rod2007,hut2014}, AFGL 4176 \citep{2015john,2020john},  Sgr C disk \citep{2022lu}, G17.64+0.16 \citep{2018maud,2019maud}, G023.01$-$00.41 \citep{2019sanna}, and G19.01-0.03 \citep{will2022}. Dedicated searches for circumstellar disks around young O-type stars suggest that disk detection rates may be highly sensitive to the evolutionary stage of these young massive stars. The dusty envelopes present in the early phases of massive protostars can obscure the presence of compact disks, while more evolved massive protostars may have already dissipated these structures \citep{2017cesa}. Highly ionized accreting disks surrounding O-type protostars have also been reported; see, for example, the cases of G345.4938+01.4677 \citep{2020andres} and
G10.6-0.4 \citep{2023roberto}. 

\subsection{Bipolar molecular outflows}\label{bipolar}

Bipolar and collimated molecular outflows are also associated with massive star formation \citep{she1996,beu2002,zha2001,zha2005,frank2014, 2015maud,li2018,2025beuther}.  
These massive molecular outflows, as the flows energized by low-mass protostars, are likely produced through entrainment of the molecular material of the parental cloud by the ionized or neutral collimated jets ejected by the young massive star. These jets, energized by the accretion process that occurs in a circumstellar disk, generate internal jet working surfaces that drive outflows \citep{raga1993}. This interaction also gives rise to bow shocks, cavities, swept-up shells, and jet-induced shocks. There are also some cases where wide-angle disk winds appear to be also present; see, for an example, the object known as Orion Source I \citep{lop2020,hiro2017} or the object Cep A HW2 \citep{torre2011}. The bipolar outflows in massive protostars show 
that the mass-loss rate (\(\dot{M}\)) in a molecular outflow increases steadily with the bolometric luminosity (\(L_{\text{bol}}\)) of the driving source, spanning a range from approximately \(1 \, L_\odot\) to \(10^5 \, L_\odot\), and that the outflow energetics also increases with the bolometric luminosity \citep{she1996,beu2002}. Additionally, \cite{beu2002} found collimation factors (that are defined as $f_\mathrm{col} = b/a$, where $a$ and $b$ are the major and minor axes, respectively) for this kind of flows on average \(\geq 2\), which is similar to results for low-mass flows, and 
reported the ``Orion-type" flows, characterized by low collimation factors, are either inherently rare or predominantly associated with very massive protostars.

The high accretion rates and the presence of disks and outflows associated with young OB stars suggest that these massive stars are capable of overcoming the formation challenges previously described and that our present model for the formation of low-mass stars \citep{ashu1987} might extend to the case of massive stars. However, even though massive star formation shares similarities with the process in low-mass stars, there are significant differences, such as the formation of ionized H\,II regions for strong UV radiation, the extremely high accretion rates, and the short accretion time scales. All these differences make
the formation of massive stars a very different and unique process.

\begin{figure}[htbp]
\centering
\includegraphics[width=0.6\textwidth]{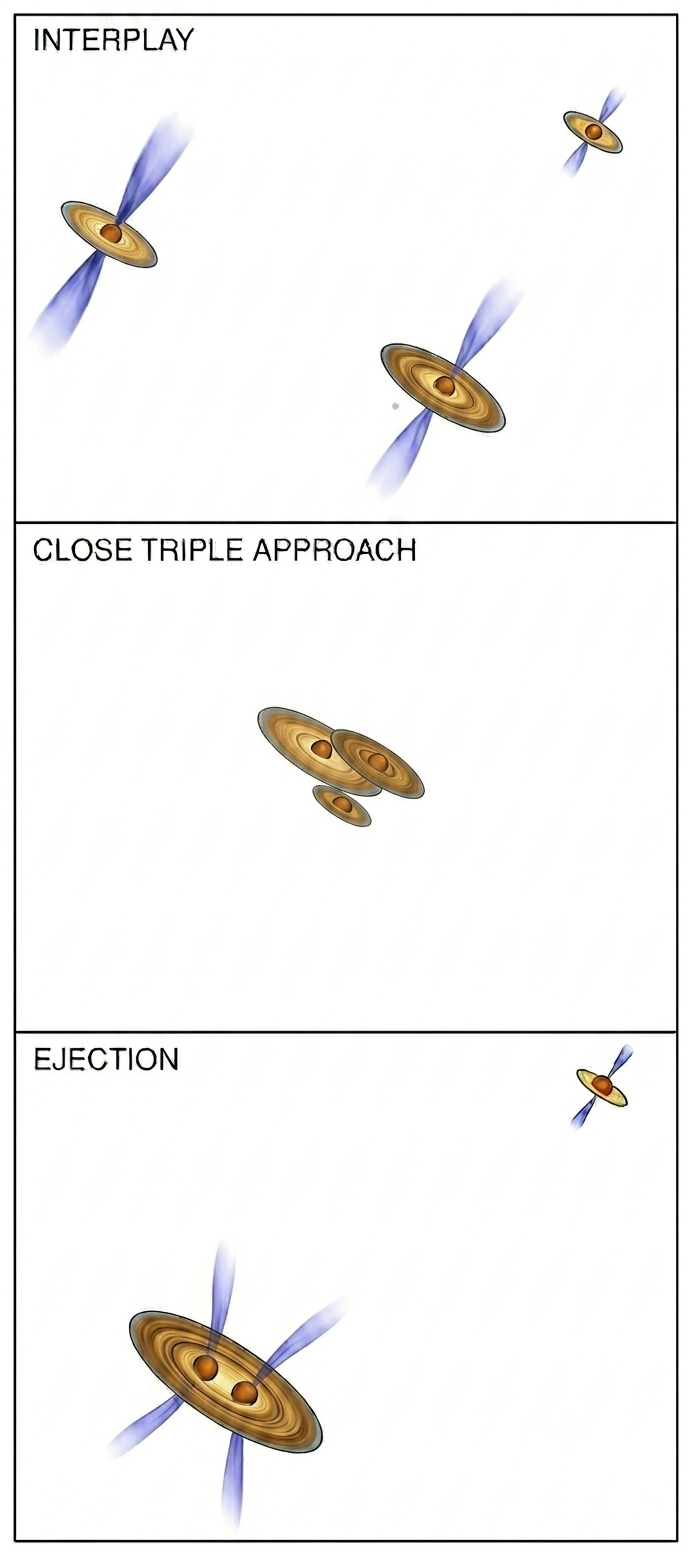}
\caption{Dynamical evolution and disintegration of a protostellar triple system.  
    \textit{Top (Interplay):} Non-hierarchical, wide-separation protostars moving chaotically, each driving independent, collimated bipolar molecular outflows from individual circumstellar disks. 
    \textit{Middle (Close Triple Approach):} Close gravitational encounter near the mutual periastron; intense tidal interactions distort and strip the individual circumstellar material. 
    \textit{Bottom (Ejection):} Gravitational slingshot results in the high-velocity ejection of the lowest-mass member (top right) into a wide orbit. The remaining two components consolidate into a tight, highly eccentric binary core (bottom left) embedded within a newly formed circumbinary disk, re-establishing tenuous, asymmetric bipolar outflows. Diagram adapted from the paradigm described by \citet{Reipurth2001}.}
\label{fig1a}
\end{figure}

\subsection{Protostellar mergers and the formation of close-by multiple stellar systems}\label{subsec2}

Given that most massive stars form in very dense young cluster environments \citep{1997hill,2005dewit}, and that some massive walkaway and runaway protostars have been observed escaping from their natal clusters \citep{hoo2000,2020dorigo,rod2020}, an additional formation mechanism for massive stars has been proposed: the merging or collision of intermediate- and low-mass protostars \citep{1998bon,2005bon,2005bally}. This scenario is closely related to dynamical interactions in dense cluster cores, where close encounters between three or four bodies can lead to the ejection of stars, the formation of tight binaries, or the merger of protostellar systems \citep{blaa1961,1967poveda}. Figure~\ref{fig1a} illustrates the schematic dynamical evolution and eventual disintegration of a non-hierarchical protostellar triple system, culminating in the formation of a tight binary pair or maybe a merger following the paradigm described by \citet{Reipurth2001}. 

The merging or collision model offers a potential solution to the radiation pressure problem, namely the difficulty of accreting material onto a protostar whose luminosity is sufficiently high to counteract the infall of gas from the parent molecular cloud. However, direct stellar collisions require extremely high stellar densities, on the order of $\sim10^{8}$ stars pc$^{-3}$, for mergers to occur frequently enough to form massive stars within $\lesssim 1$ Myr \citep{2005bon}. Observationally, young massive clusters typically exhibit stellar densities of $\sim10^{5}$–$10^{6}$ stars pc$^{-3}$ \citep{riv2013,2014riv,aina2013}, which are lower than those required for frequent direct collisions. If we translate these stellar densities now per 1000$^{-3}$ au$^{-3}$ instead of pc$^{-3}$, which are more typical scales related with radio, infrared and optical observations of proto-clusters, one can find that the stellar densities are 11.3 stars per 1000$^{-3}$ au$^{-3}$. This number does not seem difficult to reach locally in massive star-forming regions. For example, \citet{aina2018} found in OMC1 South that if one assumes that the compact mm sources will become protostars, this region will be have a high stellar densities ($\geq$ 3 stars per 1000 au$^{3}$). A second case is found in the massive star-forming region NGC~6334I(n), where \citet{li2025} reported $>$7 compact millimeter continuum sources within a region of approximately 1000~au$^3$ that exhibits a Keplerian rotation profile.

Nevertheless, the presence of extended viscous envelopes and circumstellar disks, with sizes up to several thousand astronomical units, can significantly enhance the probability of interactions and mergers among intermediate- and low-mass protostars in the dense, gas-rich centers of young clusters. In fact, the stellar densities required to perturb a binary system and induce a merger are estimated to be $\sim10^{6}$ stars pc$^{-3}$ \citep{2005bon}. As proposed by \citet{2005bon}, a binary system with a semi-major axis of a few astronomical units can be efficiently perturbed by a close encounter with a third star. The corresponding encounter radius is much larger than that required for a direct stellar collision, substantially relaxing the need for ultra-dense clusters. For example, in the case of an eccentric binary with a semi-major axis of 1 au, a stellar density of $\sim10^{6}$ stars pc$^{-3}$ is sufficient for another star to pass within an encounter radius of $\sim2$ au --- approximately the apastron separation --- within a timescale of $10^{6}$ yr \citep{2005bon}.

Once a wide binary forms through capture, it can evolve into a more massive and tighter system via continued accretion. Such close binaries are expected to eventually merge as a result of subsequent dynamical interactions with other cluster members, thereby providing a pathway for the formation of high-mass stars. This mechanism predicts that the most massive stars are preferentially single if they originate from the merger of a close binary system \citep{2005bon}. Finally, \citet{bon2003} showed that the collapse of a fragmenting core with a density of $n(\mathrm{H}_2)=10^{5}~   \mathrm{cm}^{-3}$ can transiently produce stellar densities of $10^{6}$–$10^{8}~\mathrm{pc}^{-3}$, during which stellar mergers and disk-truncating encounters are common. In their simulations, approximately one-third of all stars --- and nearly all massive stars --- experience encounters with periastron separations below 100 au. 

One of the strong barriers to form close-by binaries or mergers is the relatively high kinematic energies
of the stars in the intraclusters. If during the interaction large massive disks are disrupted, chunks
of the disks can absorb the star's kinetic energy, facilitating the formation of a binary system. Following \citet{2005bally}, one can estimate the minimum velocity ($v_m$) that a protostar must have in order to become gravitationally bound in a binary system. A star of mass $m$ can form a binary with another star of mass $M$, surrounded by a disk of mass $M_d$, if the intruder passes within the disk radius and the disk mass is sufficient to dissipate enough kinetic energy to reduce the intruder’s velocity below the escape speed of the system. This can be written as follows.
\begin{equation}
\frac{1}{2}mv_m^2<\frac{GM_dM}{r_d},    
\end{equation}
where r$_d$ is the effective half-mass radius of the disk. In this way, the surrounding disk matter-assisted capture is possible if the encounter velocity is sufficiently low. Thus, re-writing the above equation, 
one have:
\begin{equation}
v_m < \left (\frac{2GM}{r_d} \right )^{1/2} \left( \frac{M_d}{m} \right )^{1/2} < v_\mathrm{esc}(r_d) \left (\frac{M_d}{m} \right )^{1/2},    
\end{equation}
where $v_{esc}$(r$_d$) is the escape velocity at the disk radius. Let us consider a protostar of mass $M = 10\,M_\odot$ surrounded by a disk of mass $M_d = 0.1\,M_\odot$ and radius $r_d = 10$ au, and an intruding star of mass $m = 1\,M_\odot$. In this case, the intruder can be captured if its relative velocity is $\lesssim 6$\kms and if its closest approach occurs within the disk radius, allowing sufficient dissipation to reduce its velocity below the escape speed of the system.

One can also estimate the potential energy produced by the merger or the formation of a close-by system of massive stars. The energy released by a merger or the formation of a close-by binary with one of their members with a mass $m$ and the other one with a mass of $M$, is given by:
\begin{equation}
E=\frac{\epsilon GmM}{R},    
\end{equation}
where $R$ is the distance of the merger product and $\epsilon$ is an extra term that takes into account the density distributions of the stars and the mass of circumstellar material. Consider the merging of a star with mass $m$ $=$ 15 M$_\odot$ and the other one with mass $M$ = 10 M$_\odot$ at a separation of $R$ $=$ 1 au, one obtains an energy of 2.65 $\times$ 10$^{48}$ erg. 
Here, $\epsilon$ is taken to be equal to one.

In a young massive cluster there are four primary categories of recognizable bodies: disks, protostars, proto-stellar cores, and pre-stellar cores. Disks refer to the rotating structures of gas and dust surrounding young stars, which serve as the birthplace of planets. Stars, the luminous bodies formed from collapsing clouds of gas, represent the central point in stellar evolution. Pre-stellar cores are dense regions within molecular clouds where gravitational collapse is about to give rise to a new star, making them critical to understanding the early stages of massive star formation. \citet{2005bally} discussed six types of close-by interactions in young massive clusters. These interactions include then six types: core-core, core-disk, core-star, disk-disk, disk-star, and star-star.  Star-star interactions, though rare, are the most energetic and have the shortest dynamical timescales. In contrast, core-core interactions are the least energetic, but may occur more frequently due to their extensive impact areas. Intermediate in both energy and likelihood are disk-star and disk-disk interactions, which are relatively energetic and tend to occur in dense environments.

Close encounters and mergers may severely disturb and dismantle circumstellar disks.
 In the classic case of an isolated and peaceful accreting protostar, the young star is
always surrounded by a Keplerian accretion disk (with dimensions of some hundred au) and a powerful collimated outflow that reaches distances of several parsecs \citep{cod2014}.
The disk is expected to persist for several million years until erosion by photoevaporation occurs \citep{gor2009}, assuming that no external intruders interfere.

If the circumstellar disk is perturbed, the accretion activity might increase dramatically, and possibly an eruptive outflow is thus launched. Eruptive outflows in star formation refer to episodic, burst-like ejections of gas from young stellar objects, driven by sudden increases in the mass accretion rate from the circumstellar disk onto the protostar. Rather than a steady, continuous jet or wind, these systems experience instabilities in the disk (e.g., gravitational or magneto-rotational) that trigger accretion bursts, during which the accretion rate can increase by orders of magnitude.  Observationally, eruptive outflows are characterized by their clumpy or knotty structure, often seen as chains of discrete ejection events, multiple velocity components in molecular lines such as CO, and shock-excited emission produced as the expelled gas interacts with the surrounding medium. They are also closely linked to episodic accretion phenomena such as FU Orionis- and EX Lupi-type outbursts\citep{Hartmann1996,Audard2014}, highlighting the intrinsically time-variable nature of protostellar growth. During an outburst, the protostar undergoes an intensified phase of mass accretion, where material from the surrounding disk rapidly spirals onto the young star. This process releases significant energy, driving powerful jets and winds that propel gas and dust outward at high velocities. In some cases eruptive outflows or powerful accretion bursts have been reported in star-forming regions \citep{bla2025}. \citet{car2017} reported a powerful accretion burst associated with the S255IR NIRS 3 region of a high mass star formation region. However, the energetics of the outflow suggests that this outflow is perhaps produced by disk fragmentation rather than by the merging of a massive protostar (20 M$_\odot$) with a brown dwarf (0.1 M$_\odot$) with a liberated kinematic energy of $\sim$10$^{47}$ erg. The reported energy
for this object is much less than this quantity. The second case of an eruptive outflow is present in the object SVS13, an active molecular outflow emanating from a close binary source called VLA4 A and B \citep{ang2004, bla2025}. A circumbinary disk with notable spiral arms appears to feed the eruptive outflow located in this region \citep{diaz2022}. Finally, the XZ Tau outflow has also been proposed as an eruptive one, the morphology is characterized by multiple concentric structures resembling explosions or hot bubbles suggesting dynamic, episodic processes shaping the surrounding environment \citep{kri2008}. This outflow could also be driven by the interaction between a binary system of compact disks known as UZ Tau A/B \citep{kri2008,zap2015,ich2021}.    

\section{Explosive molecular outflows}\label{sec2}

\subsection{Orion BN/KL}

\begin{figure}[!ht]
\centering
\includegraphics[width=1.0\textwidth]{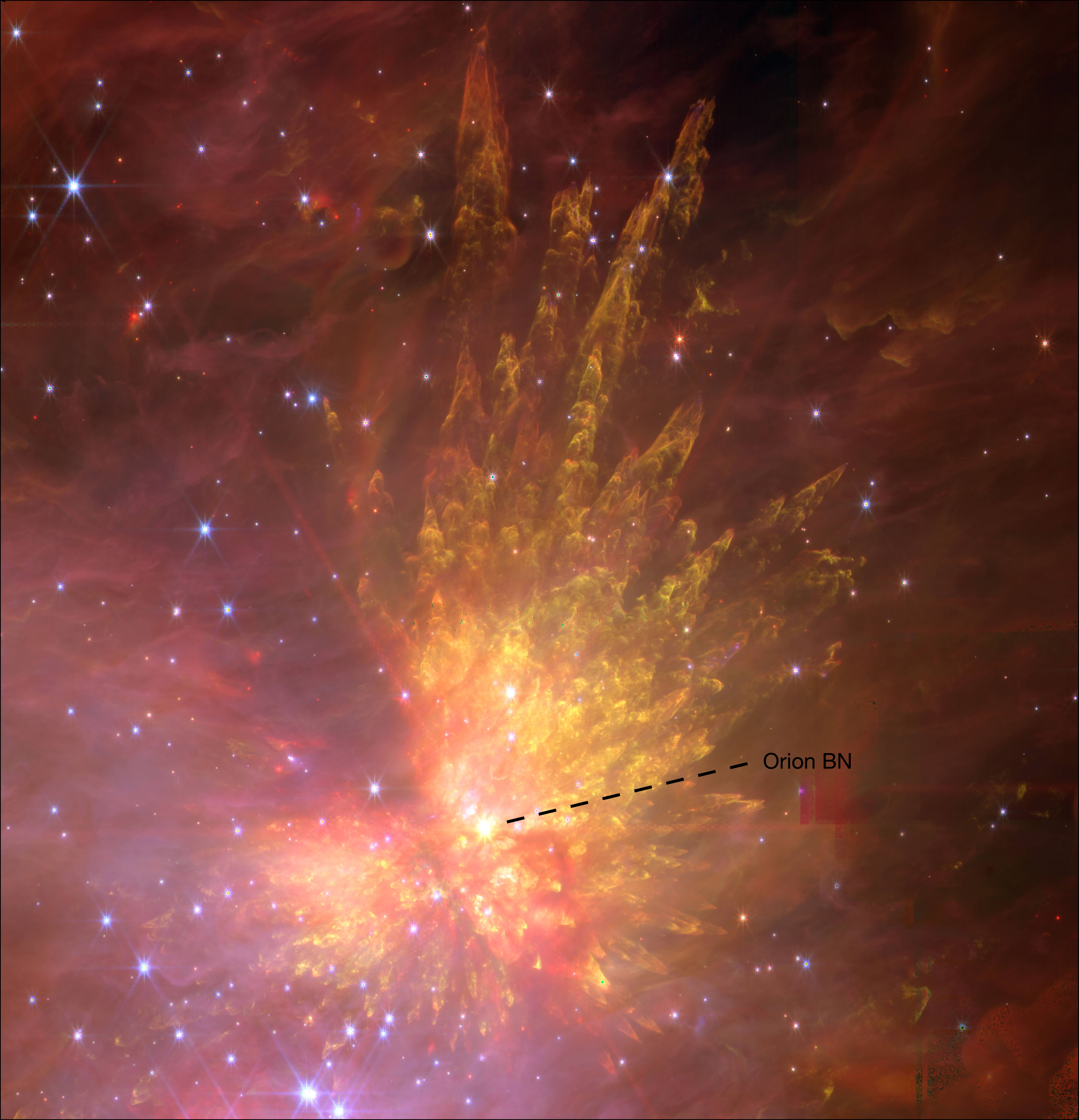}
\caption{The Kleinmann–Low Nebula (KL nebula) and the Becklin--Neugebauer Object (BN object) obtained with the James Webb Space Telescope (JWST). The Orion BN object is observed as a bright star down to the center from the image. The color composite image includes F162M (contains iron [Fe II] line), F212N (molecular hydrogen) and F470N (also molecular hydrogen). The explosive outflow in the BN/KL region is observed mainly in H$_2$ with some violet or green tips indicating the [Fe II] line emission. Note that some dusty lanes crossing in a north-east and south-west orientation are obscuring some H$_2$ fingers, this gives the impression of a non-isotropic outflow. Image Credit: NASA/ESA/CSA JWST NIRCam; Mark McCaughrean \citep{mark2023}}
\label{fig1}
\end{figure}

The Orion Becklin--Neugebauer/Kleinmann-Low (BN/KL) Nebula was named after Douglas Kleinmann and Frank J. Low, who discovered the nebula in 1967 \citep{klei1967} and Eric Becklin and Gerry Neugebauer, who found the bright infrared object BN embedded in the Orion Nebula \citep{Beck1967}, see Fig.~\ref{fig1}. This region is forming massive stars with masses slightly lower than those found in the Orion Trapezium cluster, located approximately 1$'$ to the southeast. Localized at a distance of 417$\pm$7 parsecs \citep{men2007} and behind the ONC (Orion Nebula Cluster), it is considered the closest massive star-forming region. This region has a total bolometric luminosity of $\sim$ 10$^5$ L$_\odot$ and a mass of $\sim$ 10$^3$ M$_\odot$ \citep{2008odell}. However, this massive star-forming region hosts a spectacular molecular outflow that was long believed to be driven by a massive protostar at its center.

Identified about 30--40 years ago, the molecular outflow in the BN/KL region has a peculiar, widespread, multifingered morphology reminiscent of an explosive event \citep{tay1984,allen1993,goi2026}.  
The fingerlike filaments of H$_2$ emission are extending radially outward (for more than 3$'$) from a common origin, which is the center of the BN/KL nebula; see Fig.~\ref{fig1}. Proper motions of the optical emission from the H$_2$ fingers reveal plane-of-the-sky velocities that reach up to 350\kms \citep{2000lee,doi2002,odell2008}. The proper motion vector field from these observations indicates an explosive origin about $1010 \pm 140$ years ago \citep{doi2002}, assuming no deceleration. Using observations of carbon monoxide at millimeter wavelengths, \citet{kwan1976} estimated a total mass of the outflow of 10 M$_\odot$, a kinematic energy of 4 $\times$ 10$^{47}$ erg, and
radial velocities in a range of 30--100\kms, far below the motions on the plane-of-the-sky. \citet{allen1993} estimated a mass for the compact [FeII] bullets mapped at optical wavelengths on the order of 10$^{-6}$ M$_\odot$. More recent estimations using Fabry--P\'erot v$=$l$-$0 S(l) line observations of molecular hydrogen found lower masses for the bullets of 10$^{-3}$ M$_\odot$, comparable with the Jupiter mass \citep{chr1997}. The total kinematic energy of the bullets (with a number exceeding 100) is comparable with the CO estimates, 4 $\times$ 10$^{47}$ erg \citep{chr1997}. 
\citet{chr1997} also estimated that the bullets were likely produced in a short-lived event that occurred no more than $\sim$1000 years ago.

\begin{figure}[htbp]
\centerline{\includegraphics[width=0.49\textwidth,angle=-90]{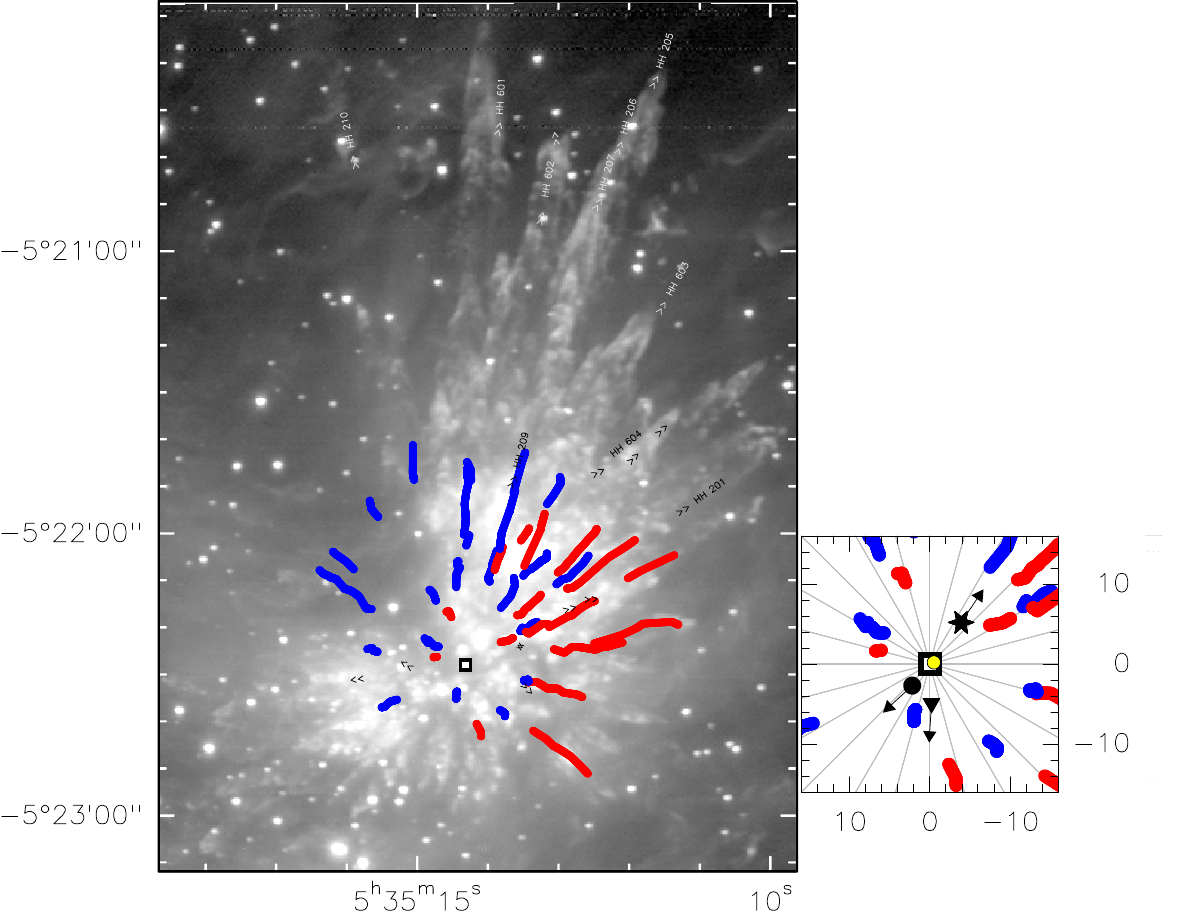}}
\vspace{3mm}
\centerline{\includegraphics[width=0.45\textwidth,angle=-90]{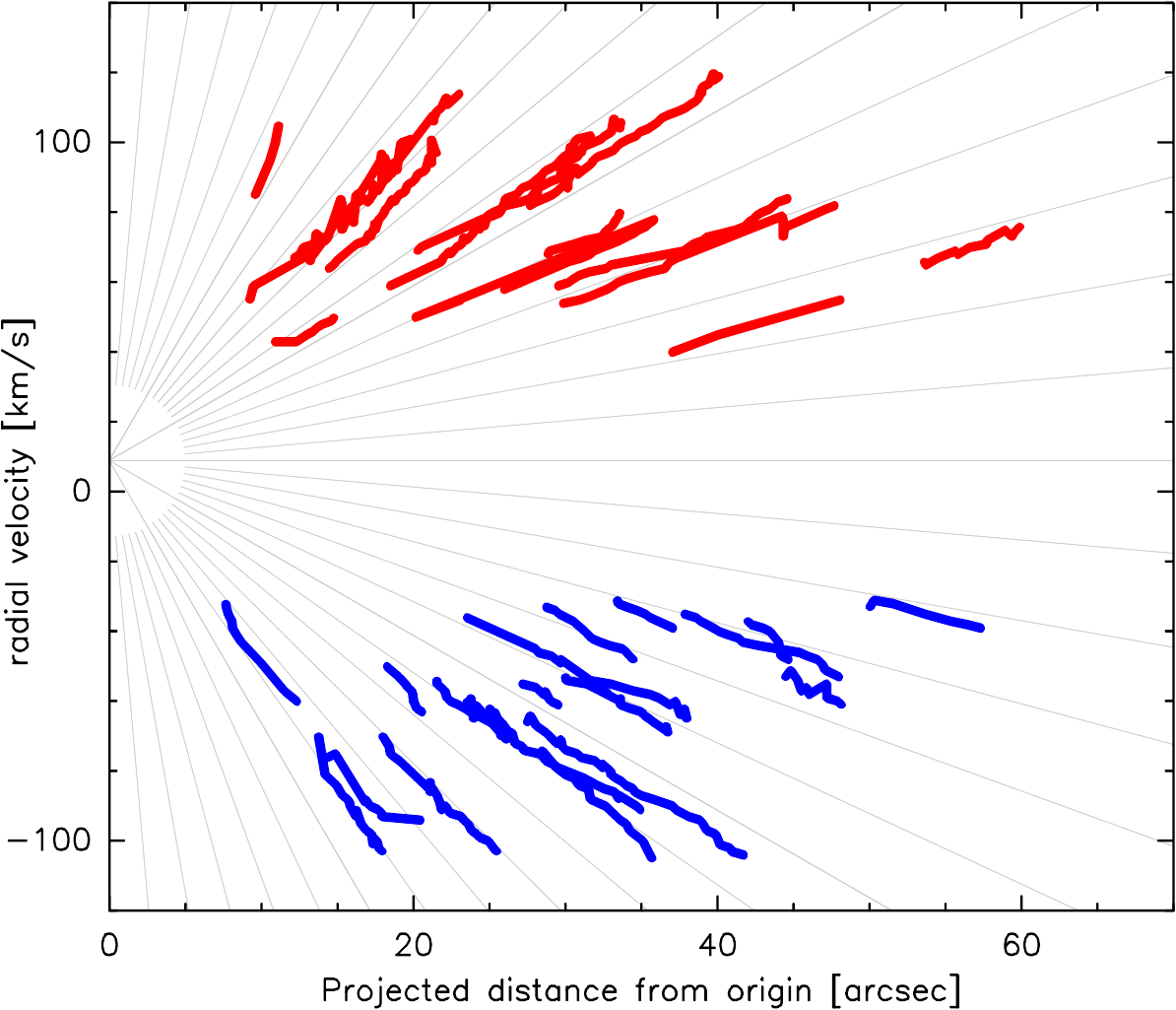}}
\caption{\textit{Upper Panel:} Blueshifted and redshifted CO(2$-$1) filaments or streamers discovered by the Submillimeter Array, overlaid on a gray scale H$_2$ image. All these molecular filaments point toward the same central position marked with a white square, and show a great correspondence with the HH objects far from the center. It is shown 
a zoom in the central region where is marked the position of the runaway young stars BN, Source I and Source n (now called Source MR, see \citealt{rod2020}). Note the coincidence of both centers, where are the molecular explosive outflow and the positions of the disintegration of the stellar system. \textit{Below:} Position-Velocity diagram from the CO(2$-$1) filaments or streamers shown above. Radial velocity as a function of projected distance from the common center for each of the 39 CO filaments, with blueshifted structures shown in blue and redshifted ones in red. All filaments appear to originate from the same common radial velocity of 9$\pm$2\kms. Image reproduced with permission from  \citet{zap2009}, copyright by AAS. }
\label{fig2}
\end{figure}

Observations of carbon monoxide at (sub)millimeter wavelengths revealed very broad wings ($>100$\kms) in its spectra, with the emission peak located close to IRc2 \citep{kwan1976,eric1982,plam1983}. Subsequent studies showed that the molecular emission spatially coincides with the H$_2$ bullets and interacts with the quiescent molecular ridge in Orion BN/KL \citep{vog1984}.

Aperture synthesis CO observations with an angular resolution of $\sim5''$ demonstrated that the outflow lobes are spatially resolved, weakly bipolar, and poorly collimated, with no evidence for a jet-like molecular component. Instead, the CO emission traces a wide ($\gtrsim 130^\circ$) biconical flow that encompasses the Herbig--Haro objects and shocked H$_2$ knots \citep{mas1984,cher1996}. Furthermore, the southeastern lobe appears to be partially weakened by the dense gas associated with the hot core \citep{cher1996}.

\begin{figure}[ht]
\centering
\includegraphics[width=0.9\textwidth]{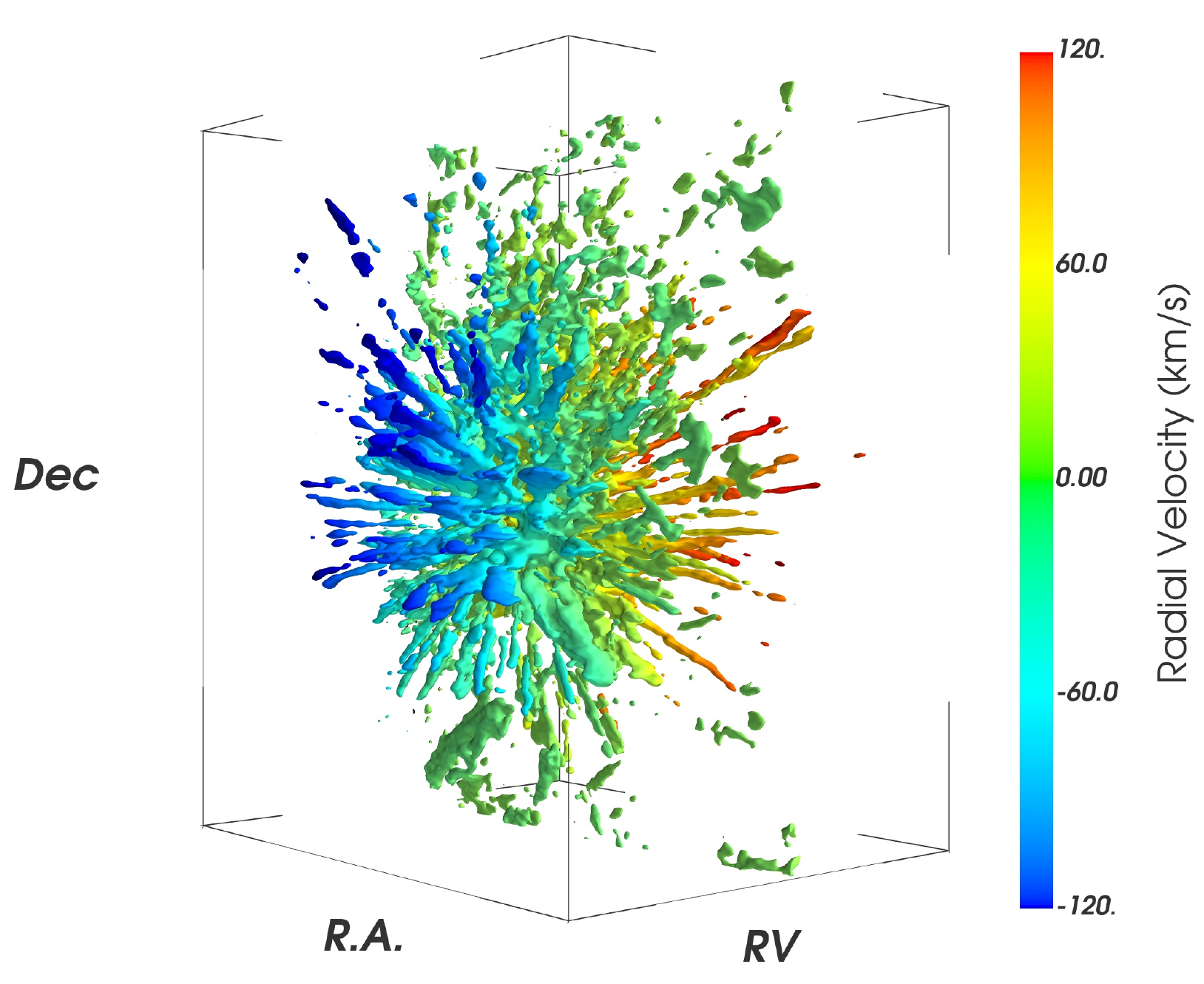}
\caption{The explosive molecular outflow in Orion BN/KL obtained with the Atacama Large Millimeter/Submillimeter Array, see \citet{bally2017}. The color image shows 3D (Right Ascension, Declination, Radial Velocity) structure from the CO(2$-$1) spectral line. The colors indicate the blueshifted and redshifted molecular emission from the explosion. The radial velocities are integrated from $-$120 to $+$120\kms. This image reveals that the outflow is mostly isotropic in all directions and shows the Hubble--Lemaître-like expansion motions. The scale-bar on the right shows the radial velocities in \kms. Image Credit: ALMA; Edgar Santamaria: Santamaria et al. (2025), in prep. }\label{fig3}
\end{figure}

\subsubsection{Confirmation}

High angular resolution (3$''$) interferometric CO(2$-$1) line observations were carried out with the Submillimeter Array\footnote{The Submillimeter Array (SMA) is a joint project between the Smithsonian Astrophysical Observatory and the Academia Sinica Institute of Astronomy and Astrophysics, and is funded by the Smithsonian Institution and the Academia Sinica. The SMA is an eight-element radio interferometer located on Maunakea, Hawaii, operating at submillimeter wavelengths (\citealt{Ho2004}). Millimeter observations with this instrument played a key role in demonstrating that an explosive event occurred in the BN/KL region.} resolved the molecular material distributed into about 40 striking jet-like streams, many of them pointing in the same direction as the H$_2$ fingers \citep{zap2009}, see Fig.~\ref{fig2}. These streamers were even mapped with single dish submillimeter observations \citep{peng2012}. The jet-like streamers 
trace nearly straight lines that converge toward a common center located between sources BN, I, and n. Their radial velocities vary linearly with the projected distance from this center, converging at $9 \pm 2$\kms, consistent with the systemic velocity of the surrounding material from BN or the hot molecular core, Fig.~\ref{fig2}. Its kinematics and morphology of the expanding stream-like filaments are reminiscent of an explosion.  This outflow therefore represents a fundamentally different class of molecular flows, distinct from the classical bipolar outflows typically produced during the process of star formation. This observation confirmed that in fact the outflow in BN/KL is explosive. Later sensitive Atacama Large Millimeter/Submillimeter (ALMA) observations confirmed the explosiveness with more than 150 striking jet-like streams, showing also Hubble--Lemaître radial velocity motions \citep{bally2017}. The Hubble--Lemaître expansion motions show velocities that increase with the distances as r$\propto$v resembling the cosmological expansion law but on much smaller scales (astrophysical).  As the CO explosive outflow is contained in a radius of about 50$''$ and taking an average radial velocity of 100\kms, one obtains a kinematic age of 700 yrs. In Fig.~\ref{fig3}, a 3-dimensional map (radial velocity, Dec, RA) is shown from the explosive outflow. The map is color-encoded with radial velocities from the CO(2$-$1) line, the blue colors represent the blueshifted velocities and the red colors represent redshifted velocities, respectively. This figure also reveals very well the
Hubble--Lemaître expansive motions in every filament, with the blue and red colors beginning to be more intense at the tips and green close to the center of the explosion. Variations in the observed angles of jet-like streams are attributable to projection effects in the plane of the sky, as noted in Fig.~\ref{fig2}. The morphology of the outflow is clearly isotropic, as revealed in the previous SMA observations.

\begin{figure}[htbp]
\centering
\includegraphics[width=0.75\textwidth, angle=0]{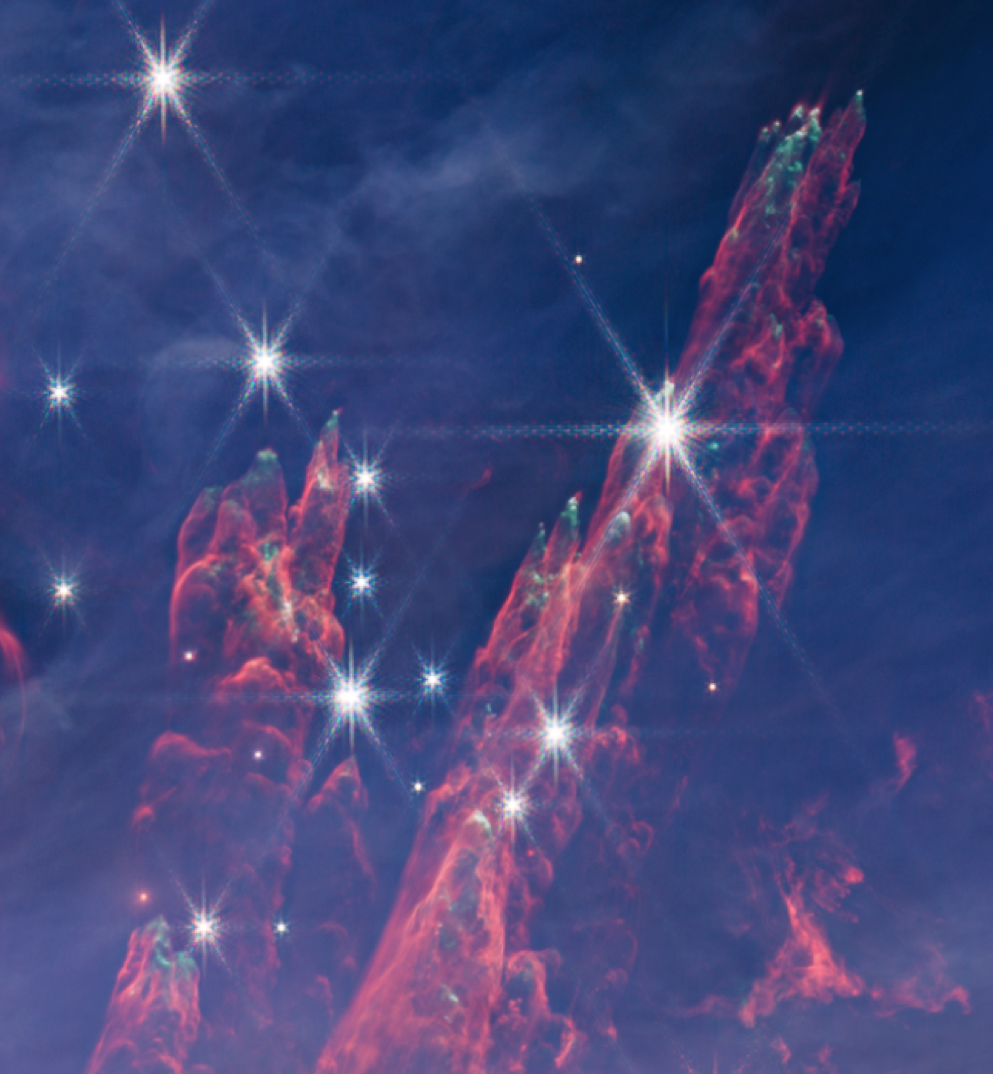}
\caption{A zoom into the tips of the fingers from the explosive molecular outflow in Orion BN/KL obtained with the James Webb Space telescope, see \citet{mark2023}. The color composite image includes F162M (contains iron [Fe II] line), F212N (molecular hydrogen) and F470N (also molecular hydrogen). The explosive outflow in the BN/KL region is observed mainly in H$_2$ (in red) with some green or violet tips indicating the [Fe II] line emission. The image clearly reveals the ``bullets'' reported for the first time in \cite{allen1993} that were ejected from the center of BN/KL. The white spots are likely foreground stars captured by the JWST.  Image Credit: NASA/ESA/CSA JWST NIRCam.}
\label{fig4}
\end{figure}

In Fig.~\ref{fig4} we show a zoom in on the tips of the infrared H$_2$ fingers captured by the JWST. This image reveals the very well-structured filament. The two filaments broke into several small filaments with compact bullets on their tips. The transversal motions for the tips have been estimated with speeds that exceed 300\kms \citep{2000lee,doi2002,odell2008,ball2011}. These infrared and optical bullets are not seen at millimeter wavelengths using molecules such as the CO, probably because the temperatures are high enough to dissociate the molecules. 
\citet{rod2023} found that to reproduce the observed morphology and kinematics of the BN/KL CO streamers and H$_2$ fingers, the clumps must first travel through a dense molecular core and subsequently emerge into a lower-density environment. This is consistent with the optical images.

\subsubsection{Search for the powering source}

At some point, the best candidate to energize the powerful outflow from
Orion BN/KL was the bright infrared source named IRc2 localized close to the center of the BN/KL outflow \citep{fra1999,gen1989,gen1981,dow1981}. This is based on the classical view that a massive star is exciting a bipolar molecular outflow, as observed in other regions \citep{torre2011,hiro2017,bally2023A}. However, high angular infrared observations quickly showed that this source is multiple, casting doubt on the possibility that a single very luminous source ($\sim$ 10$^5$ L$_\odot$) is responsible for the energetics in the region \citep{dou1993,gen1992}. Several compact radio and infrared objects were subsequently identified as potential exciting sources in the vicinity of the outflow center; among them are the radio Source I, Orion BN, Orion Source X, and Radio Source n. 

\paragraph{Radio Source I:} Source I is a deeply embedded massive young stellar object with an estimated mass of 10--20 M$_\odot$ \citep{Reid2007,Goddi2011} and with high proper motions 12$\pm$3\kms \citep{Rod2005,lau2005,rod2020}. More recent molecular observations have placed this source at a dynamical mass of 15$\pm$2 M$_\odot$ \citep{manon2018}. This source drives a bipolar molecular outflow oriented north-east/south-west, with high-velocity shocks probed by H$_2$O, OH and SiO masers \citep{gau1998,gree1998,kim2008} and thermal SiO emission \citep{beu2005,zap2012,lop2020,wri2024}.   
High-resolution VLA and ALMA studies show a compact source elongated roughly along the disk plane, with emission that rises smoothly from centimeter to millimeter wavelengths without the turnover expected from a classical H II region \citep{rei2008}. Instead of being dominated by free–free radiation from an ionized nebula, the spectrum is consistent with partially ionized, dense gas in a disk wind, possibly produced by collisions or shocks within the outflowing material \citep{pla2017,wri2023}. Recently, in this object, sodium chloride (NaCl), potassium chloride (KCl) and their isotopologues ($^{37}$Cl and $^{41}$K) have been found, molecules present mainly in evolved stellar ejecta \citep{mamon2019}. 

\begin{figure}[ht]
\centering
\includegraphics[width=0.75\textwidth]{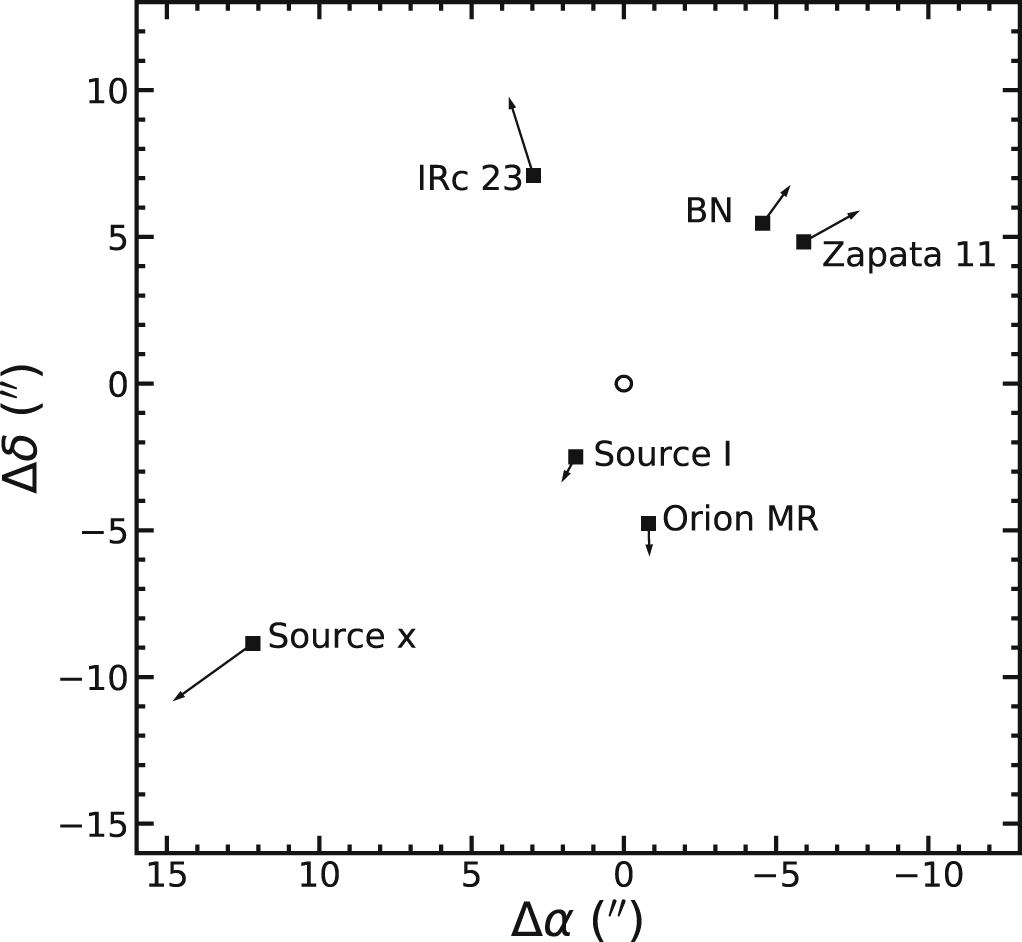}
\caption{Proper motions (arrows) and locations (squares) of the six objects (Source I, IRc 23, BN, Zapata 11, Orion MR, and Source X) reported with large proper motions in the center of Orion BN/KL region. The size of the arrows marks the proper motion in 100 yrs. The direction of the arrows represents the orientation of the proper motions. The empty circle shows the estimated center of the explosive outflow in BN/KL. Image reproduced with permission from \citet{rod2020}, copyright by AAS. }
\label{fig5}
\end{figure}

\paragraph{Orion BN:} The Becklin--Neugebauer object (BN) is a massive young stellar object in the Orion BN/KL region, with an estimated mass of $\sim$ 8--15\,M$_\odot$ \citep{Beck1967,Sco1983}. Unlike Source I, BN is directly observable in the infrared regime, where it appears as one of the brightest compact infrared sources in the Orion Nebula. In Fig.~\ref{fig1}, it is observed
to the northwest as one compact and bright source, just above the dusty and dark ridge.
BN exhibits compact radio continuum emission, interpreted as arising from a partially ionized stellar wind or hypercompact H II region \citep{plam1995,Rod2005}. BN also presents a large proper motion in the plane of the sky, with a magnitude of 30\kms and an orientation to the northwest \citep{Rod2005,lau2005,rod2020} from the center of the explosive outflow in BN/KL, see Fig.~\ref{fig5}.

\paragraph{Orion Source X:} Source X is a compact infrared object localized in the vicinities of Orion BK/KL. This was
identified in infrared surveys with Keck and VLT as a faint reddened object \citep{Luh2017}. \citet{Luh2017} estimated a mass for this object of 2--3\,M$_\odot$, a bolometric luminosity of 20--30 L$_\odot$, and an effective
temperature of 5000 K. \citet{Bally2020} confirmed that the luminosity is not as high as BN or Source I, confirming that this source is more likely a low- or intermediate-mass young star. The proper motions estimated for this source are about 55\kms with an orientation to the southeast from the center of the explosive outflow in BN/KL \citep{Luh2017,Bally2020}.

\paragraph{Orion Source n:} Orion Source n has been detected in the infrared and as a compact radio continuum source, consistent with a partially ionized stellar wind \citep{men2007,Rod2005}. Infrared interferometry and imaging reveal an elongated structure interpreted as a dusty circumstellar disk, oriented nearly perpendicular to a bipolar outflow traced in shocked H$_2$ and radio emission \citep{Shu2004}. Proper motion measurements show that Source n is moving southward at about 15\kms, again from the center of the explosive outflow in BN/KL \citep{lau2005}. Recently, \citet{rod2020} presented a reanalysis of the proper motions in the Orion BN/KL region using an extended set of VLA observations and showed that the radio source previously associated with the infrared source n is in fact a distinct object with significant southward motion, which they designate as Orion MR while the infrared counterpart (Source n) remains essentially stationary; see Fig.~\ref{fig2}.  
In addition, they identified other compact radio sources with large proper motions, including IRc23 and Zapata 11, that, together with BN, Source I, Orion MR, and Source X, appear to be moving away from a common point at the center of the outflow in BN/KL. On the other hand, \citet{Bally2020}, based on near-infrared 2.2 $\mu$m imaging, proposed that IRc23 and Zapata 11 might not be runaway stars, but features associated with a slow bipolar outflow driven by BN.

\begin{figure}[ht]
\centering
\includegraphics[width=\textwidth]{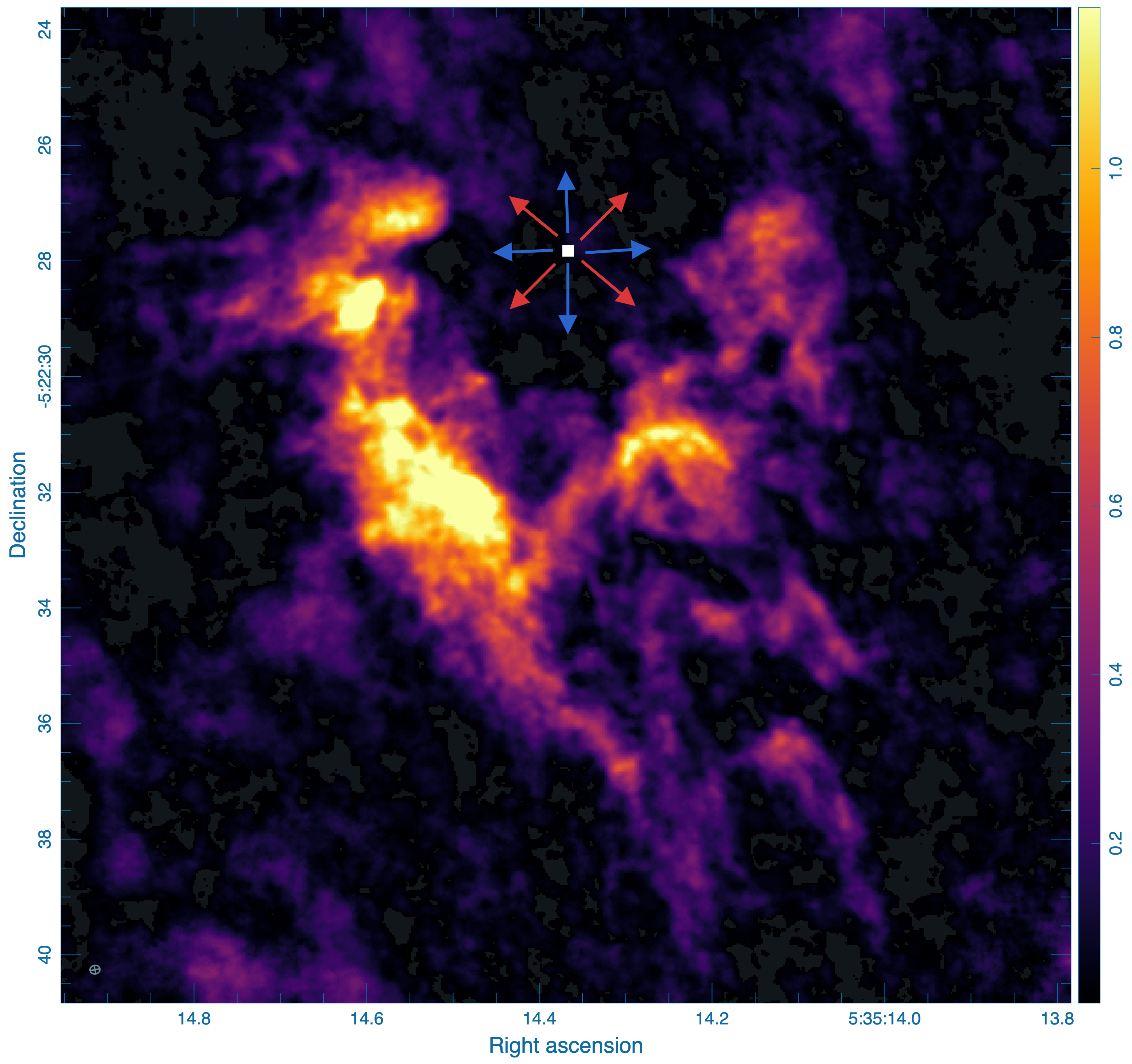}
\caption{ALMA moment zero or integrated intensity map from the Ethyl Cyanide (CH$_3$CH$_2$CN[N=25$-$24 J=24$-$24] at a rest frequency of 232.79002 GHz) thermal emission from the Orion Hot Molecular Core. The ALMA data is obtained from project 2016.1.00165.S, P.I. John Bally. The map was computed in a velocity range from $-$5 to $+$25\kms. The white square marks the position of the center from explosive outflow shown in Fig.~\ref{fig2} and ~\ref{fig3}. The red and blue arrows trace the streamers from the explosion, some of them are impacting in a dense region creating the hot molecular core.  There is a clearly deficiency of streamers where is located the hot core, indicating that some of them have impacted the dusty ridge in Orion creating the Orion Hot Core, see Fig.~\ref{fig7}.}
\label{fig6}
\end{figure}

\begin{figure}[!h]
\centering
\includegraphics[width=0.95\textwidth]{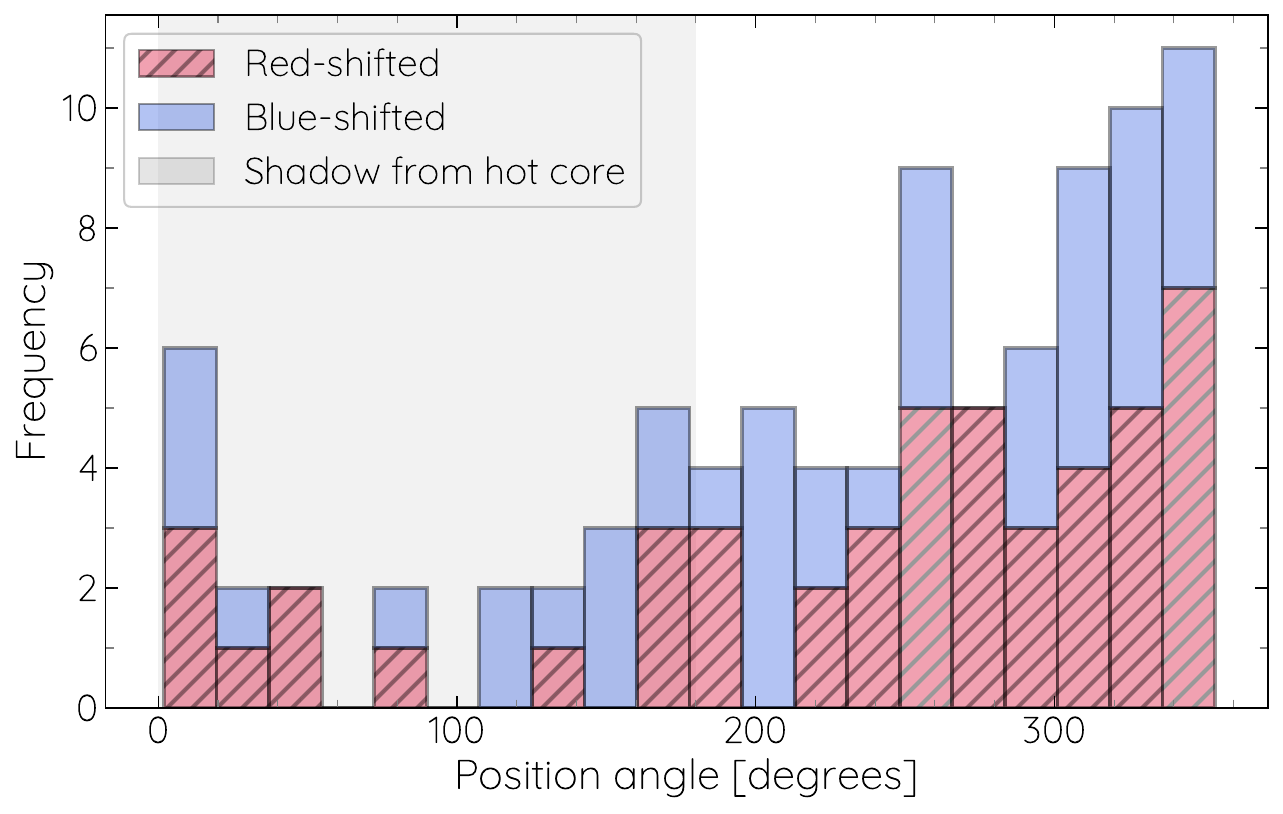}
\caption{Histogram from the blue-shifted and redshifted molecular streamers in Orion BN/KL obtained with ALMA data. The plot shows the frequency vs. the position angle in the plane-of-sky for all streamers shown in Fig.~\ref{fig3}. We have also included position angles where is located the shadow the hot molecular core in Orion, similar to \citet{zapata2011} made with data from the Submillimeter Array. Here, there is a clear deficiency of streamers at the location of the hot core, indicating that some of them have impacted the dusty ridge in Orion creating the Orion hot core. This image is obtained from Santamaria et al. (2026). in prep.}
\label{fig7}
\end{figure}

As can be seen in Fig.~\ref{fig5}, all runaway infrared, radio, and optical sources appear to converge to the estimated origin of the explosive molecular outflow reported in \citet{zap2009}. This indicates that both phenomena are closely related, the ejection of the young stars and the molecular outflow in BN/KL \citep{zap2017}. Therefore, this outflow constitutes a separate class of molecular flows, unlike the classical bipolar outflows of forming stars. The most plausible scenario is that the stellar system disintegrated about 500 yr ago \citep{lau2005,rod2020}, launching the outflow at approximately the same dynamical time. The estimated kinematic age of the outflow, about 700 yr \citep{zap2009}, is consistent with the timescale of the stellar disintegration. Initially, Source I, Orion n (now Orion MR), and Orion BN were found to exhibit large proper motions, with their trajectories converging to a common position about 500 years ago. This strongly suggested that they were once members of a compact massive multiple system that underwent dynamical decay \citep{lau2005,2005bally}. In this scenario, Source I has been proposed to be a close binary system (of a few au) or a merger formed during the decay. Probably, at some point, four massive young stars interacted dynamically and formed the merger/binary system, and after that the three objects were then ejected from the center of BN/KL. The gravitational potential energy released during the merger is sufficient to account for the kinetic energy of the explosive outflow, which is on the order of $\sim10^{48}$~erg. However, given the relatively large number of runaway sources (see Fig.~\ref{fig5}), the classic three-body interaction scenario is unlikely, since it would produce only two escaping bodies, a compact binary, and the third star of the encounter \citep{rod2020}.  A new model needs to be proposed that includes all massive runaway stars, the explosion, and the formation of the molecular shell or bubble that surrounds the explosion center \citep{zapata2011}, and finally the formation of the bright hot molecular core that will be discussed in detail below. This molecular bubble is composed of gas and dust that probably were once part of the circumstellar environment surrounding the now disrupted multiple system and now are expanding at a radial velocity of 15\kms, a velocity similar to those of young massive stars \citep{zapata2011}.

\subsubsection{The hot molecular core}

Since its discovery through hot thermal NH$_3$ emission \citep{ho1979}, the physical nature of the hot molecular core in Orion BN/KL (Fig.~\ref{fig6}) has been the subject of extensive discussion \citep{blake1996, liu2002, cher1996}.  Recent observational evidence suggests that the heating of this core is primarily driven by explosive streamers that originate in the BN/KL region \citep{zapata2011,tere2017,pag2017,pag2019}. Figure~\ref{fig1} shows that several molecular H$_2$ streamers appear to penetrate the dense dust ridge, extending down to the bright and compact BN star.  In Fig.~\ref{fig7}, on the other hand, a histogram of the number of molecular streamers is present (see Fig.~\ref{fig3}) as a function of position angle. The distribution shows a decrease in the number of streamers toward the hot molecular core, suggesting that many of them are blocked by a dense cloud located within the dusty ridge. This result clearly indicates that the explosive outflow is primarily responsible for heating the hot core, as also reported in \citet{zapata2011}. In this scenario, the southeastern and southwestern sectors of the outflow have collided with a pre-existing, very dense portion of the extended ridge, generating shocks that produced the bright Orion KL hot core. Some other more recent studies have also suggested external heating in the Orion hot molecular core, which includes \citet[however, they suggested the outflow from Source I as exciting source for the hot core]{god2011,wri2017}, \citet{fav2011,peng2012,peng2013,bell2014,gon2015,peng2017,li2020}.

\citet{fav2017,pag2017,pag2019} using sensitive ALMA observations to the Orion BN/KL hot core have revealed how the explosive outflow impacts several zones of the dusty ridge and many complex organic molecules are formed on the icy surfaces of dust grains and are subsequently released into the gas phase through CO-desorption with water, carried by shocks of matter (steamers) ejected during the explosive event. They also proposed that the Compact Ridge is sufficiently distant from the rest of the BN–KL hot core to have remained unperturbed by the explosion. This interpretation is supported by the absence of asymmetric emission-line wings and by the narrow width of the lines themselves ($\sim$1\kms).  
Finally, there is recent evidence of internal heating in some millimeter continuum cores within the Orion BN/KL region\citep{wil2022}. 

Figure~\ref{fig6} presents an ALMA map of ethyl cyanide emission toward the Orion Hot Core, overlaid with the position of the explosion center and the orientation of several molecular streamers. The molecular emission is well resolved spatially, revealing a rich and complex internal structure. The molecular material exhibits numerous arcs and filaments that appear to point back to the origin of the explosive outflow, likely formed contemporaneously with the explosive event. Kinematic maps of the hot core molecular gas reveal a wide range of velocities among individual clumps, suggesting that the explosion accelerated the gas in multiple directions \citep{zapata2011,god2011,peng2013}. 

\clearpage
\subsection{G5.89--0.39}

\citet{har1988} reported that the molecular outflow in G5.89$-$0.39 (W28A2) is one of the most powerful outflows in the Galaxy, with a kinetic energy of $\sim$10$^{49}$ erg. However, more recent estimates place the energetics of this notable outflow in a lower range, between $\sim$10$^{46}$ and $10^{48}$ erg \citep{acco1998,klaa2006}. Nevertheless, it remains one of the most energetic outflows associated with massive star-forming regions.
G5.89$-$0.39 has been recognized by its strong ultra compact H\,II region, which is the center of the outflow and studied in many wavelengths; see the references in \citet{2021FernandezLopez}. 
The ultra compact H\,II region has a morphology reminiscent of a ring or torus with a dynamical age of 600 yr, estimated from its expansion rate 
\citep{acco1998}. At one point, the exciting source of this ultracompact H II region was proposed to be a massive young O5 V star, identified through near-infrared observations obtained with the Nasmyth Adaptive Optics System (NAOS) and the Near-Infrared Imager and Spectrograph CONICA (NACO) at the Very Large Telescope (VLT; NACO–VLT; \citealt{feldt2003}). However, this young massive infrared source is significantly offset from the center of the ultracompact H II region by about 1$''$. One possible explanation is that the source migrated from the region with an escape velocity of $\sim$10\kms roughly 1000 years ago \citep{feldt2003}. \citet{klaa2006} proposed that the outflow in this region is a fossil flow --- that is, an outflow lacking a currently active driving source but continuing to propagate through momentum conservation --- with an estimated age of $\sim$1000 yr, similar age to that of the expanding ultracompact H\,II region. Later CO and SiO millimeter wave observations revealed a complex outflow with multiple extremely high-velocity ($-$150 to $+$80\kms) lobes with different orientations \citep{su2012,sollins2004}.  
In addition, the outflow lobes clearly exhibit a Hubble--Lemaître-like kinematic structure (as the ones observed in the explosive outflow from the BN/KL region), with velocities increasing linearly with distance from the center.

\subsubsection{Proposal and confirmation}

\citet{zapata2019} using high angular resolution (0.85$''$) CO(3$-$2) Submillimeter Array observations report the possible detection of an explosive molecular outflow in this massive star-forming region. The data reveal six narrow, high-velocity molecular streamers that were ejected from a common central position. These filaments exhibit a Hubble--Lemaître-like kinematic structure, with velocities increasing linearly with the projected distance from the apparent explosion center. The inferred dynamical age of the outflow is of the order of $\sim10^{3}$~yr, comparable to the expansion timescale of the central ultra-compact H~{\sc ii} region and the age estimated in \citet{klaa2006}. The morphology and kinematics of the flow are inconsistent with steady collimated protostellar jets.

\begin{figure}[ht]
\centering
\includegraphics[width=\textwidth]{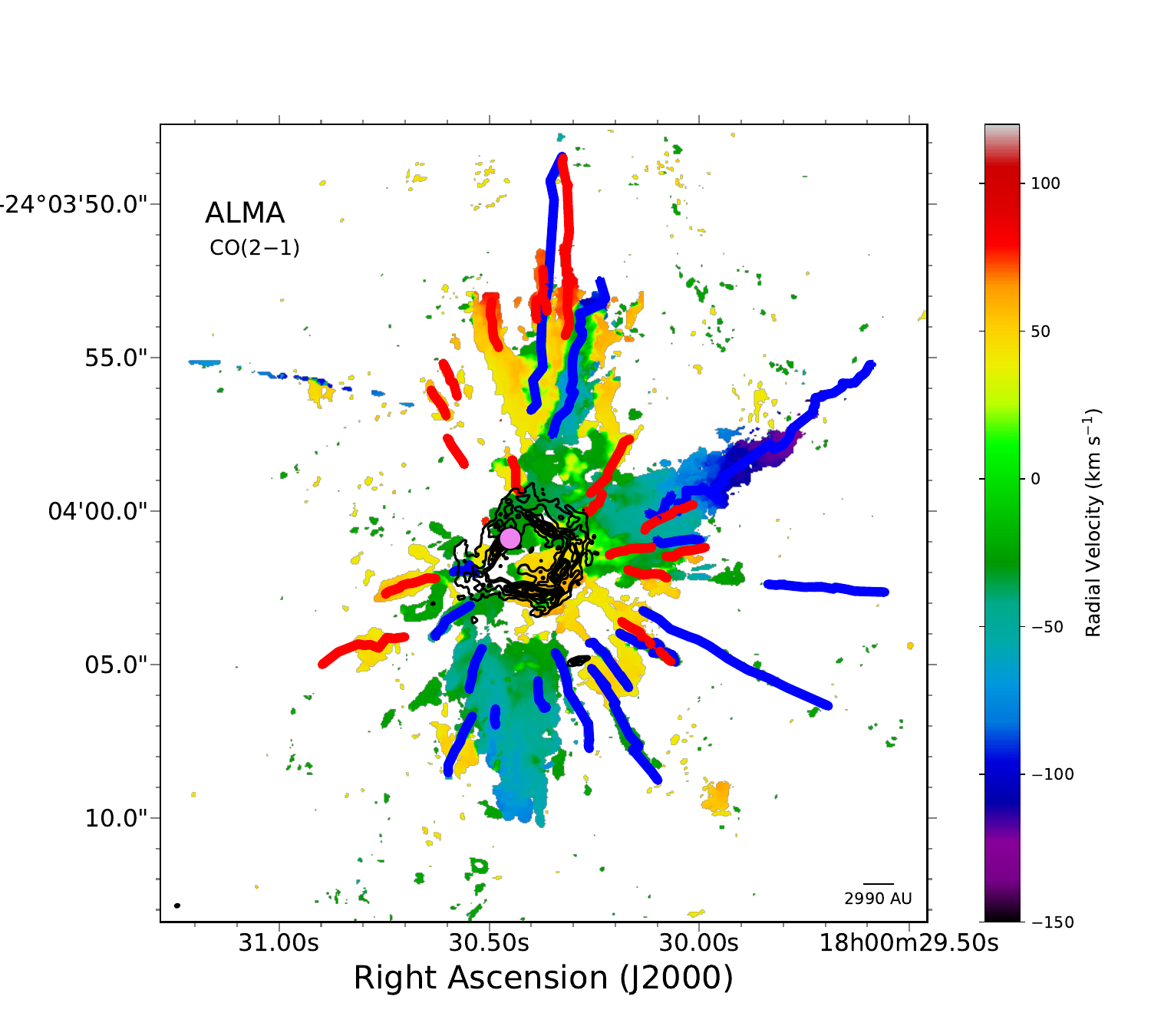}
\caption{The explosive molecular outflow (color scale) and the ultra compact H\,II region (black contours) in G5.89$-$0.39 obtained with the Atacama Large Millimeter/Submillimeter Array.  The colors indicate the blueshifted and redshifted molecular emission from the explosion. The radial velocities are integrated from $-$150 to $+$120\kms. This image reveals that the outflow is mostly isotropic in all directions and shows the Hubble--Lemaître-like expansion motions. The scale color bar on the right shows the radial velocities in \kms. The location of the source named Feldt’s star \citep[pink circle]{feldt2003} is shown at the center of the explosive outflow. The half-power contour of the synthesized beam of the line image is shown in the bottom left corner. 
Image reproduced with permission from \citet{zap2020}, copyright by AAS}
\label{fig8}
\end{figure}

\begin{figure}[ht]
\centering
\includegraphics[width=1.0\textwidth]{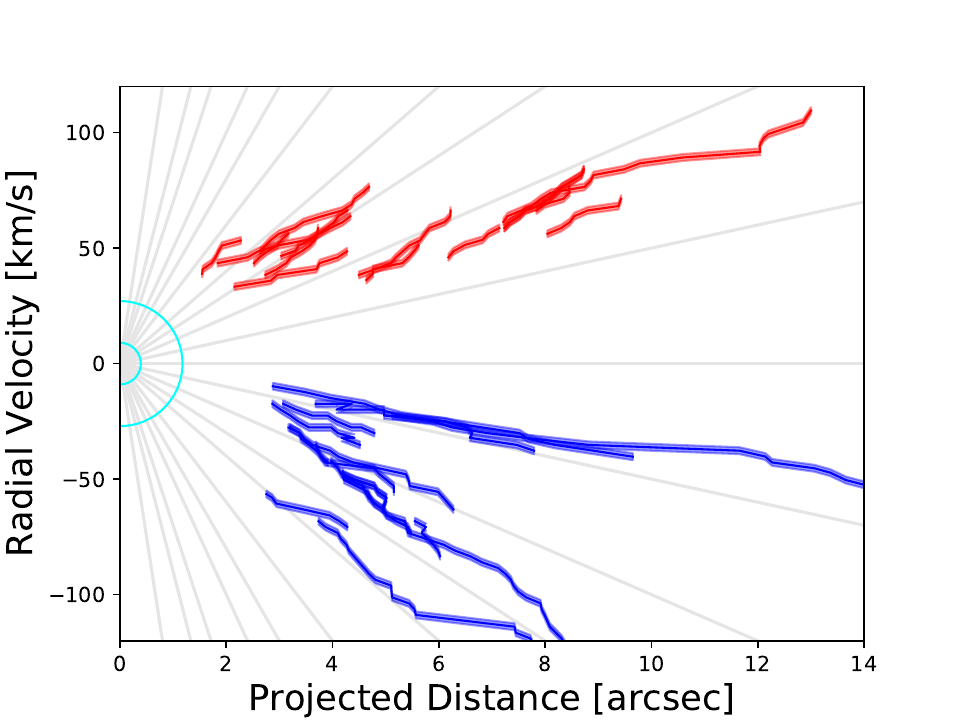}
\caption{Position--velocity diagram of the explosive CO(2--1) streamers in the G5.89$-$0.39 outflow revealed by ALMA observations. Only the maximum velocities within the range of $\pm$120\kms are shown; several molecular filaments extend beyond this velocity window. This is colored with the approaching filaments in blue, and the receding ones in red. 
Image reproduced with permission from \citet{zap2020}, copyright by AAS}
\label{fig9}
\end{figure}
  
This scenario for G5.89$-$0.39 was confirmed using sensitive and high angular resolution ALMA observations \citep{zap2020}. The ALMA data revealed more than $\sim$30 narrow molecular filaments (Figs.~\ref{fig8} and \ref{fig9}) emanating radially from a common center located near the ultra compact H\,II region, with radial velocities that increase linearly with distance from the origin, producing a clear Hubble--Lemaître-like kinematic pattern (Fig.~\ref{fig9}), as observed from the Submillimeter Array observations. 

From the observed distances and velocities, a dynamical age of $\sim$1000~yr is inferred, indicating a short-lived, impulsive event. Since the dynamical timescales of explosive outflows are very short, these phenomena are believed to originate from a single, sudden energetic event rather than from a long-term continuous process. Therefore, the physical mechanism responsible for producing such outflows must involve an impulsive release of energy occurring within a brief period of time. This dynamical age is in good correlation with that estimated before \citep{klaa2006,zapata2019}. Strong SiO emission close to the expansion center traces shocked gas, supporting the interpretation of a sudden release of energy. The estimated kinetic energy of the outflow, $\sim10^{46}$--$10^{48}$~erg, is far larger than that of typical protostellar outflows but well below supernova energies.
   
\subsubsection{Outflow driving mechanism}

From all these observations, this is then argue that the explosive molecular outflow in G5.89 is most plausibly explained by a brief, energetic event such as a (proto)stellar merger or the formation of a close binary system, similar to the scenario proposed for Orion~BN/KL \citep{bally2017}. This interpretation naturally accounts for the large velocities and the stream/filamentary morphology observed in the molecular gas.

Adopting a representative radial velocity of $\sim$80\kms for the molecular filaments at a projected distance of $\sim$6$''$ from the inferred center of the explosion, it is derived a kinematic age of order $\sim$1000~yr for the molecular outflow. This timescale is consistent with the dynamical ages estimated for both the expanding ionized shell ($\sim$600~yr) and the fossil CO outflow \citep[$\sim$1000~yr;][]{klaa2006}, suggesting that these phenomena are likely manifestations of the same underlying event.

Assuming an age of $\sim$1000~yr for the explosion, the projected separation of $\sim$1.5$''$ ($\sim$4300~au) between Feldt’s infrared source and the center of the outflow implies a plane-of-the-sky velocity of order $\sim$20\kms for the infrared star. This value is comparable to the tangential velocities measured for the runaway sources BN, Source~I, and~MR in Orion \citep{rod2007}, supporting the idea that Feldt’s star may have been dynamically involved in the event that powered the outflow. 

\citet{2007watson} reported an extended east-west, older molecular outflow emanating from this region, which could be fossil outflow similar to that reported associated with DR21 \citep{skre2023}.

\subsection{Other cases} \label{sec:other}

To date, evidence for explosive outflows has been reported in several other massive star-forming regions. In the following, a brief overview of these regions is provided.   

\subsubsection{DR21} \label{sec:dr21}

DR21 is a massive star-forming region in the Cygnus~X complex that hosts an exceptionally energetic molecular outflow ($\geq$ 10$^{48}$ erg) with an estimated mass of $\geq$ 3000~M$_\odot$. Early CO observations revealed high-velocity gas that extends on parsec scales \citep{gar1991,gar1992}, while later interferometric studies showed that the outflow is composed of multiple high-velocity filaments or streamers with wide-angle morphology \citep{Zap2013}. The combination of its morphology, energetics, and short dynamical timescale ($\sim$8600 yrs) is difficult to reconcile with a steady, single collimated jet, and instead suggests an origin in a brief, explosive event, similar to that proposed for Orion~BN/KL or G5.89 \citep{colque2024}. A second possibility is that the extended outflow in DR21 represents an older, fossil flow, while the explosive outflow is a more recent event \citep{skre2023, colque2024}. 

Using sensitive high-angular-resolution VLA observations, \citet{Yan2026} identified multiple ionized arc-shaped structures that can be fitted with parabolas, whose symmetry axes converge toward a position coincident with the proposed center of the explosive outflow, suggesting that the giant H\,II region is also related with the explosive outflow, as in the G5.89. The existence of the ionized arc-shaped structures pointing to the center of the outflow suggests that at this position still exist a massive star or stars, in contrast to the BN/KL case, where no stars have been detected at the center of the explosion.  
\subsubsection{IRAS 16076--5134} \label{sec:IRAS16076} 

IRAS 16076$-$5134 or AGAL 331.28$-$00.19 is a massive star forming region localized at 5 kpc, with a total luminosity of 2$\times$10$^5$ L$_\odot$ and that is associated with a strong UCH\,II region. IRAS\,16076$-$5134 exhibits clear signatures of an explosive outflow, as revealed by sensitive ALMA observations at 0.9\,mm and in the CO(3$-$2) line \citep{guzman2022}. The data reveal an ensemble of streamer-like CO ejections with radial velocities ranging approximately from $-62$ to $+83$\kms that appear to emanate radially from a common central position near the dense core MM8, resulting in a quasi–isotropic distribution on the plane of the sky \citep{guzman2022}. Several of these streamers show linear radial velocity gradients that increase with distance from the putative center, reminiscent of Hubble--Lemaître-like expansion motions associated with explosive events \citep{guzman2022}. The estimated dynamical age of the outflowing material is approximately 3500 yr, with a momentum of $\sim$10$^4$ M$_\odot$\kms) and a high kinetic energy (10$^{48-49}$ erg). This massive star-forming region is thus suggested to host an explosive molecular outflow with one of the largest kinetic energies known.

\subsubsection{IRAS 12326--6245} \label{sec:IRAS12326} 

IRAS\,12326$-$6245 is a luminous ($\sim$10$^5$ L$_\odot$) massive star-forming region in which recent ALMA Band~6 observations at $\sim$0.2$''$ resolution have uncovered clear evidence for an explosive molecular outflow \citep{zap2023}. The data reveal more than ten well-defined molecular streamers with Hubble--Lemaître-like expansion motions, radiating roughly isotropically from a common central position. These streamers are observed in CO emission and are highly collimated, indicating that the outflow is not a classical bipolar jet but instead has a dispersal morphology characteristic of explosive events.
A dusty and molecular shell, newly reported in this work and located in the northern part of the UC\,H\textsc{ii} region G301.1A, coincides with the apparent origin of the outflow streamers. From the observed velocities and projected distances of the streamers, it is estimated a kinematic age for the explosive event of approximately 700\,yr and a kinetic energy of order $10^{48}$\,erg, consistent with other known explosive outflows in massive star-forming regions \citep{zap2023}. 

\subsubsection{Sh2-106} \label{sec:sh2-106} 

In addition to molecular explosive outflows, evidence for short-lived, energetic feedback events has also been identified in ionized gas. A notable example is the bipolar H\,II region Sh2-106, where multi-epoch \textit{Hubble Space Telescope} observations reveal supersonic expansion of the ionized nebula, with velocities increasing with distance from the central massive source S106~IR \citep{ball2022}. The inferred expansion speeds, reaching $\sim$150\kms, and a kinematic age of $\sim$3500~yr are difficult to reconcile with steady stellar winds alone and instead point to a brief, energetic outburst. Although Sh2-106 lacks the prominent molecular streamers seen in Orion~KL or IRAS~12326$-$6245, its kinematics and energetics suggest that explosive feedback may manifest across different gas phases, from molecular to fully ionized material. Taken together, these sources indicate that episodic, high-energy events associated with massive young stellar objects may be a common, yet short-lived, phenomenon during early cluster evolution \citep{ball2022}.

\subsubsection{G34.26+0.15} \label{sec:g34.26} 

Recent ALMA archival observations reveal evidence for a likely explosive outflow in the high-mass star-forming complex G34.26+0.15 \citep{isa2025}. Combining CO(2–1) and SiO(5–4) data, the authors identify multiple high-velocity molecular streamers that originate from a common center embedded within an ultracompact H\,II region. These streamers exhibit radial velocities spanning roughly 0–120\,km\,s$^{-1}$, and their kinematics follow a Hubble--Lemaître-like velocity law, characteristic of explosive dispersal outflows rather than steady bipolar jets. From the CO emission, the total outflow mass is estimated to be $\sim$264\,M$_\odot$, with a momentum of $\sim4.3\times10^3\,$M$_\odot$\,km\,s$^{-1}$ and a kinetic energy of $\sim10^{48}$\,erg, placing the event among the most energetic explosive outflows observed in massive star formation. The inferred dynamical age of the outflow is approximately $1.9\times10^4\,$yr, which is comparable to the age of the associated ultracompact H\,II region, suggesting a possible causal connection between the explosive event and the ionized gas expansion. Interestingly, the orientation of the magnetic field in the region appears broadly aligned with the outflow streamers, indicating that the explosion may have influenced the local magnetization as well \citep{isa2025}. 

\subsubsection{IRAS 15520–5234} \label{sec:IRAS 15520}

\citet{hoque2026} reported the presence of an explosive outflow in the massive protocluster IRAS 15520$-$5234 (IRAS 15520), based on high-angular-resolution ALMA Band 6 observations obtained as part of the QUARKS survey. Their analysis revealed compelling evidence for an explosive molecular outflow characterized by 28 highly collimated “fingers” or “streamers” that follow a Hubble--Lemaître-type velocity law and emanate from a common center. The event released a kinetic energy of at least $4.1 \times 10^{48}$ erg, substantially exceeding the typical energetics of bipolar outflows associated with massive protostars. The dynamical age of the flow is estimated to be $\sim 6.6 \times 10^{3}$ yr. Based on the current census of known explosive outflows, they estimate that such events occur in the Milky Way at a rate of approximately one per $\sim 80$ year. They suggested that mass redistribution and dynamical interactions among massive cores within the protocluster likely triggered the explosive event. Further high-angular-resolution observations are needed to constrain the underlying driving mechanism and confirm its origin.

\begin{landscape}
\begin{table*}
%%\centering
\caption{Physical properties of massive star-forming regions hosting explosive outflows}
\label{tab:explosive_outflows_refs_foot}
\begin{tabular}{lccccccc}
\toprule
Source & $L_\mathrm{bol}$ & $M_\mathrm{gas}$ & Tracer & $\Delta v$ & Age & $E_\mathrm{kin}$ & $\dot{E}$ \\
 & ($L_\odot$) & ($M_\odot$) &  & (km\,s$^{-1}$) & (yr) & (erg) & (erg\,s$^{-1}$) \\
\midrule

Orion BN/KL 
& $\sim10^{5}$ 
& $\sim10$ 
& CO, H$_2$, [Fe\,II] 
& 10--150 
& 500--700 
& $10^{47}$--$10^{48}$ 
& $10^{37}$--$10^{38}$ \\

DR21 
& $10^{5}$--$10^{6}$ 
& $10^{2}$--$10^{3}$ 
& CO 
& 10--100 
& $\sim8.6\times10^{3}$ 
& $\sim10^{48}$ 
& $\sim10^{36}$ \\

G5.89$-$0.39 
& $3\times10^{5}$ 
& $\sim10^{2}$ 
& CO, SiO 
& 20--160 
& $\sim10^{3}$ 
& $10^{46}$--$10^{48}$ 
& $\sim10^{37}$ \\

IRAS 16076$-$5134 
& $\sim10^{5}$ 
& 100--200 
& CO 
& 20--80 
& $\sim3.5\times10^{3}$ 
& $10^{48}$--$10^{49}$ 
& $10^{36}$--$10^{37}$ \\

IRAS 12326$-$6245 
& $4\times10^{5}$ 
& 50--100 
& CO 
& 20--100 
& $\sim7\times10^{2}$ 
& $10^{47}$--$10^{49}$ 
& $\sim10^{37}$ \\

G34.26+0.15 
& $\sim10^{6}$ 
& 200--300 
& CO, SiO 
& 20--120 
& $\sim2\times10^{4}$ 
& $\sim10^{48}$ 
& $10^{35}$--$10^{36}$ \\

Sh2\,106 
& $\sim10^{5}$ 
& 10--50 
& Ionized gas 
& 30--150 
& $\sim3.5\times10^{3}$ 
& $\sim10^{47}$ 
& $\sim10^{36}$ \\

IRAS 15520$-$5234 
& 4 $\times$ 10$^{5}$ 
& 24
& CO 
& 10--80 
& $\sim6.6\times10^{3}$ 
& $\ge4.1\times10^{48}$ 
& $\sim2\times10^{37}$ \\

\bottomrule
\end{tabular}
%%\tablefoot
{\scriptsize

$L_\mathrm{bol}$ is the bolometric luminosity.
$M_\mathrm{gas}$ is the approximate gas mass associated with the explosive outflow.
$\Delta v$ indicates the observed velocity range.
Ages are dynamical estimates and should be considered order-of-magnitude values.
$E_\mathrm{kin}$ is the kinetic energy of the accelerated gas.
$\dot{E} = E_\mathrm{kin}/t_\mathrm{dyn}$ is the average mechanical luminosity.
All quantities are subject to projection effects, optical depth effects, and tracer-dependent uncertainties.
References: Orion BN/KL \citep{ball2011,zap2009,bally2017};
DR21 \citep{gar1991,colque2024};
G5.89$-$0.39 \citep{klaa2006,su2012,sollins2004,zap2020};
IRAS 16076$-$5134 \citep{guzman2022};
IRAS 12326$-$6245 \citep{zap2023};
G34.26+0.15 \citep{isa2025};
Sh2\,106 \citep{ball2022};
IRAS 15520$-$5234 \citep{hoque2026}.
}
\end{table*}
\end{landscape}

\subsection{Common properties}\label{sec:common}

Table~\ref{tab:explosive_outflows_refs_foot} summarizes the main physical properties of massive star-forming regions that host explosive outflows, including molecular and ionized examples. Despite the diversity of environments, several common characteristics emerge.

\begin{itemize}
    
\item  First, all regions are associated with high bolometric luminosities, typically $L_\mathrm{bol} \gtrsim 10^{5}\,L_\odot$, indicating that explosive outflows preferentially occur in sites of massive star formation. This suggests that the presence of massive protostars or young O-type stars is a necessary condition to power such energetic events. In contrast, classical low-mass bipolar outflows are rarely associated with comparable luminosities or energetics and are powered continuously by jets detected as faint thermal radio sources \citep{ang2018}.

\item  Second, these regions contain substantial gas reservoirs, with characteristic gas masses ranging from $\sim10$ to a few $10^{2}\,M_\odot$, and reaching values of order $10^{3}\,M_\odot$ in particularly massive complexes such as DR21 and G34.26+0.15. The availability of large amounts of dense molecular material likely plays a key role in shaping the observed explosive morphology, as it provides both the mass accelerated during the event and the medium that records the radial streamer-like kinematics.

\item  A striking common feature of explosive outflows is the exceptionally wide velocity range exhibited by the molecular streamers, with $\Delta v$ spanning from several tens up to $\sim150$\kms. Such broad velocity distributions are difficult to produce through steady driving and instead strongly favor an impulsive acceleration mechanism.

\item In all cases, the kinematics are characterized by Hubble--Lemaître-like expansion, in which the flow velocity increases approximately linearly with distance from the explosion center; this behavior, consistently reported in the literature, is a hallmark of ballistic expansion and provides strong support for an explosive origin.

\item  The inferred dynamical ages are relatively short, typically $\sim10^{3}$~yr, with the exception of DR21 and G34.26$+$0.15, which exhibit older ages of order $10^{4}$~yr. These longer timescales may indicate either older explosive events or the superposition of multiple episodes of energetic feedback. In this context, DR21 has been proposed to host both a fossil large-scale outflow and a more recent explosive event, suggesting that explosive phenomena may recur during the early evolution of massive stellar systems.

\item  Taken together, the combination of high luminosity, large gas mass, wide velocity range, and short dynamical timescale strongly favors scenarios involving dynamical interactions in young massive stellar systems, such as (proto)stellar mergers or the formation and hardening of close multiple systems. The similarity of these properties across different regions, including the ionized case of Sh2-106, suggests that explosive outflows may represent a distinct and relatively common mode of feedback in massive star formation rather than rare or exceptional events.

\end{itemize}

\clearpage
\subsection{Mechanical vs. bolometric luminosity}\label{sec:bolo} 

A direct comparison between the explosive outflows listed in
Table~\ref{tab:explosive_outflows_refs_foot} and classical protostellar outflows reveals fundamental differences in both energetics and temporal behavior. Typical molecular outflows driven by steady disk-mediated accretion in massive protostars exhibit kinetic energies of
$E_\mathrm{kin}\sim10^{43}$--$10^{45}$~erg and characteristic lifetimes of
$\sim10^{4}$--$10^{5}$~yr, corresponding to energy injection rates, or mechanical luminosities, of $L_\mathrm{mech}\equiv\dot{E}\sim10^{33}$--$10^{35}$~erg~s$^{-1}$ \citep[e.g.,][]{arce2007,frank2014}.
In contrast, the explosive outflows summarized here reach kinetic energies of $E_\mathrm{kin}\sim10^{47}$--$10^{49}$~erg over significantly shorter dynamical timescales ($\sim10^{3}$--$10^{4}$~yr), implying instantaneous energy injection rates, or mechanical luminosities, of
$L_\mathrm{mech}\equiv\dot{E}\sim10^{35}$--$10^{38}$~erg~s$^{-1}$, see Table~\ref{tab:explosive_outflows_refs_foot}.
These values exceed those of classical outflows by one to three orders of
magnitude, even when conservative assumptions on gas mass, velocity, and
inclination are adopted. Such extreme energetics are difficult to reconcile with continuous jet or wind models and instead point toward impulsive events, such as stellar dynamical interactions or mergers, as the dominant powering mechanism.

\begin{figure}[ht]
\centering
\includegraphics[width=1.0\linewidth]{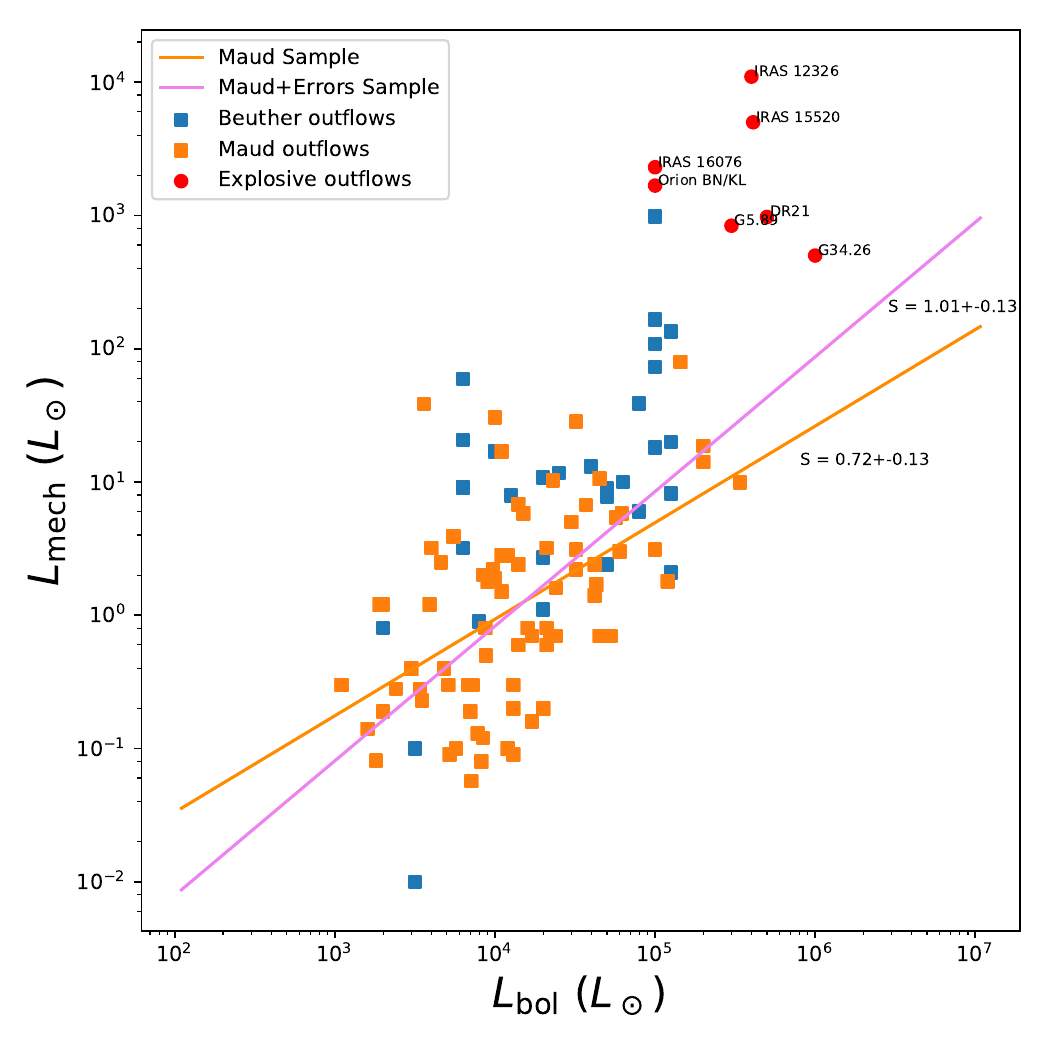}
\caption{
Mechanical luminosity versus bolometric luminosity.
Blue and orange squares represent classical molecular outflows compiled from the samples
of \citet{beu2002} and \citet{2015maud}, together with the empirical scaling 
relation $L_\mathrm{mech}\propto L_\mathrm{bol}^{0.69\pm0.12}$ from \citet{2015maud}. Red circles indicate explosive outflows, including Orion~BN/KL, G5.89$-$0.39, DR21, IRAS~16076$-$5134, IRAS~12326$-$6245, IRAS 15520$-$5234, and G34.26$+$0.15.
Explosive outflows lie systematically far above the classical trend.
}
\label{fig:fig10}
\end{figure}

Figure~\ref{fig:fig10} presents the mechanical luminosity ($L_\mathrm{mech}\equiv\dot{E}$) as a function of bolometric luminosity for massive and luminous protostars. Classical molecular outflows, shown as blue and orange squares from the samples of \citet{beu2002} and \citet{2015maud}, follow the well-known empirical scaling relation $L_\mathrm{mech}\propto L_\mathrm{bol}^{0.6-0.7}$ \citep{wu2004,2015maud}, consistent with steady, accretion-driven feedback. In this figure, we include the correlation derived for the Luke T. Maud sample (orange line). We also include the same sample after adding artificial uncertainties of 0.15 dex to the bolometric luminosity and 0.3 dex to the mechanical luminosity, which yields a relation of $L_\mathrm{mech} \propto L_\mathrm{bol}^{1.1}$ (pink line). In stark contrast, explosive outflows --- highlighted by red circles and including Orion~BN/KL, G5.89$-$0.39, DR21, IRAS~16076$-$5134, IRAS~12326$-$6245, IRAS 15520$-$5234, and G34.26$+$0.15 --- systematically lie far above this relation. For a given bolometric luminosity, their mechanical luminosities exceed those of classical massive protostellar outflows by one to two orders of magnitude, underscoring again the fundamentally different, impulsive nature of their driving mechanism.

\begin{figure*}[ht!]
\centering
\includegraphics[width=1.0\linewidth]{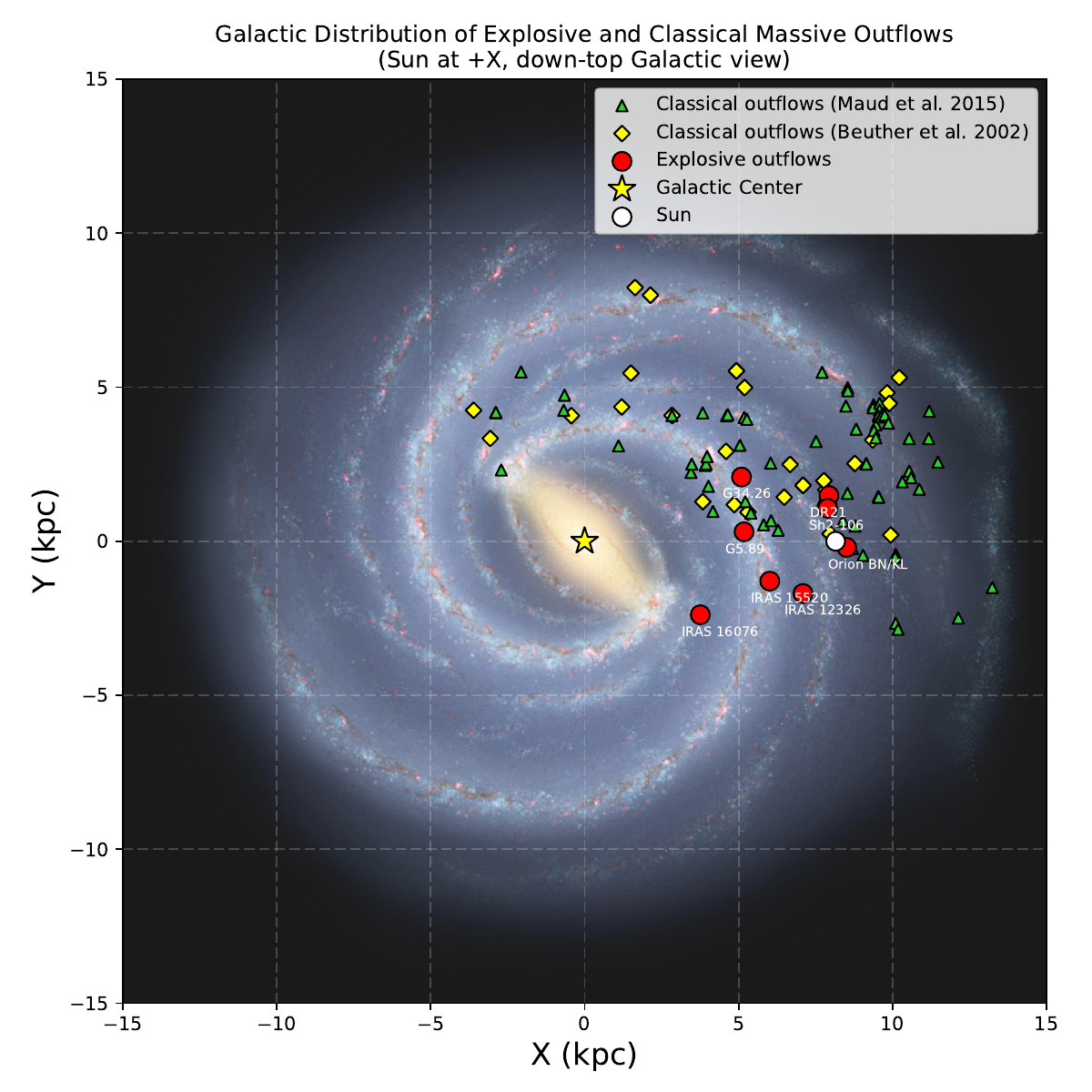}
\caption{
Projected Galactocentric distribution of massive protostellar outflows from \citet{2015maud} (green triangles) and \citep{beu2002} (yellow diamonds), and regions hosting explosive outflows (red circles), shown in a top-down view of the Milky Way. The Galactic Center is located at the origin of the coordinate system, and the Sun is placed at $(R_0,0)$ with $R_0 = 8.15$~kpc. Individual heliocentric distances are obtained from the Maud et al.\ and Beuther et al.\  samples, allowing the large-scale spatial distribution of classical massive protostellar outflows to be compared directly with that of explosive outflows. The background image provides a schematic representation of the Galactic disk and is included for visual reference only.}
\label{fig:fig11}
\end{figure*}

\newpage
\subsection{Galactic distribution and event rate} \label{sec:distri}

The projected Galactocentric distribution of regions hosting explosive outflows shows that these events are not confined to a specific Galactic environment. Instead, they are observed across a wide range of Galactocentric radii and Galactic longitudes. As illustrated in Fig.~\ref{fig:fig11}, explosive outflows are found both in the inner Galaxy and along the outer spiral arms, spanning distances from the Galactic center comparable to those of classical massive protostellar outflows. This broad spatial distribution indicates that the physical conditions required to produce explosive outflows---such as the presence of massive protostars and high stellar densities --- are not exclusive to extreme environments near the Galactic center, but can also be achieved in more typical massive star-forming regions throughout the Galactic disk.

Moreover, the absence of any obvious spatial clustering of explosive outflows supports the idea that they arise from localized, stochastic dynamical processes within individual young massive clusters, rather than being regulated by large-scale Galactic conditions.

An estimate of the Galactic rate of explosive outflows can be obtained by considering the dynamical ages and distances of the eight objects listed in Table~\ref{tab:explosive_outflows_refs_foot}. Neglecting projection effects on the velocities, we note that seven explosive outflows are observed within a timescale of $\sim8{,}600$~yr (the kinematic age of DR21) and within a radius of 5.0~kpc (the distance to IRAS~16076$-$5134). The source G34.26$+$0.15 is excluded from this estimate because its explosive outflow dynamical age is more uncertain \citep{isa2025}. This yields an inferred rate of approximately one explosive event every $\sim1{,}200$~yr within this area. Scaling this rate to the full Galactic disk, assuming a Galactic radius of $1.4\times10^{4}$~pc \citep{Rix2013,Bland-Hawthorn2016}  -- measured for the stellar and molecular thin disk: 

\begin{equation}
\frac{r_E^2}{r_G^2} \simeq \frac{25}{196}.
\end{equation}
where $r_E$ is the radius of the area that produces explosive flows and $r_G$ is the Galactic radius, resulting in an estimated Galactic rate of roughly one explosive flow every $\sim150$~year.

An independent estimate can be obtained using a steady-state argument that relates the number of observed events to their characteristic visibility timescale. If $N_\mathrm{EO}$ is the number of known explosive outflows in the Galaxy and $\tau_\mathrm{EO}$ is their typical dynamical age, the Galactic occurrence rate $\mathcal{R}_\mathrm{EO}$ can be approximated as
\begin{equation}
\mathcal{R}_\mathrm{EO} \simeq \frac{N_\mathrm{EO}}{\tau_\mathrm{EO}}.
\end{equation}

Taking the explosive outflows listed in Table~\ref{tab:explosive_outflows_refs_foot} (Orion~BN/KL, DR21, G5.89$-$0.39, IRAS~16076$-$5134, IRAS~12326$-$6245, G34.26$+$0.15, IRAS~15520, and Sh2\,106), and conservatively adopting $N_\mathrm{EO}\sim8$ and characteristic dynamical ages of $\tau_\mathrm{EO}\sim10^{3}$~yr, we obtain an estimated Galactic explosive event rate of

\begin{equation}
\mathcal{R}_\mathrm{EO} \sim 8\times10^{-3}~{\rm yr}^{-1},
\end{equation}
corresponding to roughly one event every $\sim125$~yr. This value is remarkably similar to the previous estimate and to those reported in earlier studies \citep{zapata2019,zap2020,colque2024}.

Following \citet{zap2020}, this rate can be compared to the formation rate of massive stars ($\gtrsim8\,M_\odot$). The global star formation rate of the Milky Way is $\sim2\,M_\odot\,{\rm yr}^{-1}$ \citep{eli2022}. Assuming an average stellar mass of $\sim0.5\,M_\odot$ for the initial mass function \citep{par2011}, this corresponds to the formation of approximately four stars per year. Given that massive stars constitute only $\sim0.5\%$ of the stellar population \citep{par2011}, the implied massive star formation rate is roughly one star every $\sim50$~yr, a value comparable to the Galactic supernova rate \citep{tam1994}. The inferred explosive outflow rate is within an order of magnitude of the massive star formation rate, implying a physical link between these phenomena and suggesting that a significant fraction of massive stars undergo an explosive outflow phase during their formation.

The current census of explosive outflows is certainly preliminary as their identification requires high-angular-resolution, high-sensitivity molecular line observations capable of resolving faint, high-velocity streamers. Such observations are strongly biased toward nearby, luminous regions that have been targeted with interferometers, while more distant or older explosive events may have faded below detectability or been misclassified as more conventional bipolar outflows.

\begin{figure*}[t!]
\centering
\includegraphics[width=1.0\linewidth]{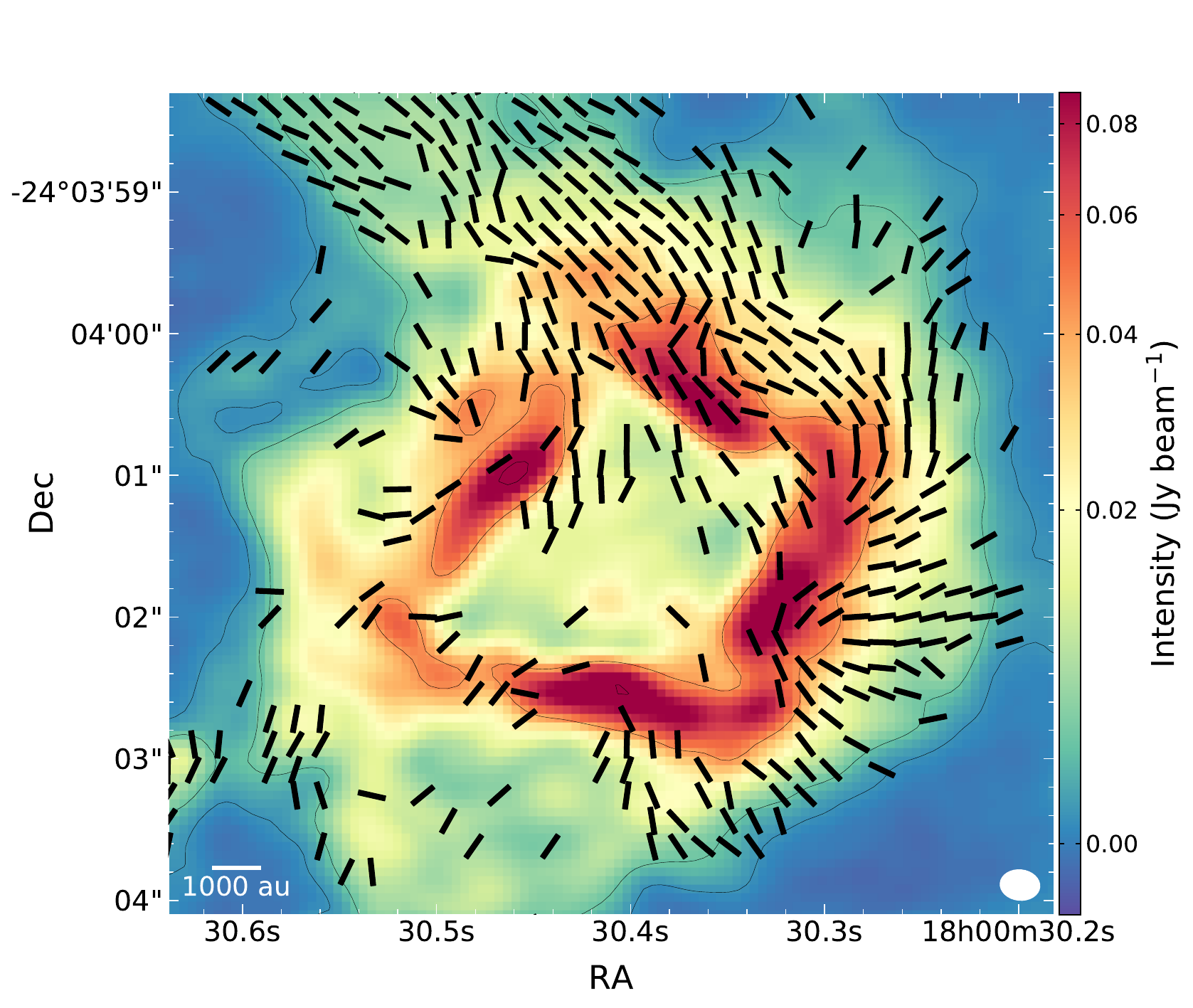}
\caption{Stokes I continuum image of toward the G5.89-0.39 region (colour-scale) overlayed with magnetic field orientation sticks \citep[black segments][]{2021FernandezLopez}. The center of the EO is located at the center of the dust and ionized-gas shell. The B-field morphology within the shell is more azimuthal, whereas outside the shell it is more radial.}
\label{fig:regions}
\end{figure*}

\subsection{Magnetic fields} \label{sec:bfields}

Millimeter polarization observations have been reported toward at least three explosive outflows known to date. In all cases, the dominant mechanism responsible for the detected linear polarization has been interpreted as thermal emission from dust grains aligned with the magnetic field \citep[e.g.,][]{1994Lazarian,2015Andersson}. While this assumption is generally valid at (sub)millimeter wavelengths, the extreme physical conditions associated with explosive outflows may also allow for alternative polarization mechanisms to operate.

Early BIMA observations toward Orion~BN/KL suggested that specific localized regions might be dominated by mechanically aligned grains \citep{1998Rao}, as originally proposed by \citet{1952Gold}. More recent high-angular-resolution ALMA observations have further shown that intense radiation fields from massive protostars --- in particular Orion~Source~I --- propagating through the outflow cavities may also contribute to grain alignment and produce detectable polarized emission \citep{2021Pattle}. Nevertheless, the prevailing interpretation of millimeter linear polarization remains magnetic alignment, and in the following we discuss the available observations under this assumption. To date, such studies exist for Orion~BN/KL \citep{2010Tang,2017Pattle,2021Cortes}, G5.89$-$0.39 \citep{2021FernandezLopez}, and G34.26$+$0.15 \citep{2024Khan,isa2025}.

A striking and common result among these three regions is the presence of an overall radial magnetic field configuration associated with the explosive outflows. This morphology suggests a connection between the magnetic field structure and the quasi-isotropic distribution of the molecular streamers, possibly indicating that the field lines have been dragged outward by the expanding outflow under conditions of magnetic flux-freezing. Although significant local irregularities are present in all cases, the radial pattern extends to radii as small as $\sim$1{,}500~au from the inferred explosion centers.

In detail, G34.26$+$0.15 exhibits a radial magnetic field pattern primarily in its northwestern half \citep{2024Khan}. In Orion~BN/KL, the field shows considerable irregularity between $\sim$1{,}500 and 5{,}000~au \citep{2021Pattle}, transitioning to a quasi-hourglass morphology at larger scales \citep{2017Pattle,2021Cortes}. In G5.89$-$0.39, the magnetic field displays a patchy, predominantly azimuthal configuration at radii of $\sim$4{,}500~au, coincident with a dense ring of dust and gas, while becoming preferentially radial beyond this distance \citep{2021FernandezLopez}. This behavior is consistent with a flux-freezing scenario in which magnetic field lines are compressed within dense condensations of the expanding ring and are dragged outward where the ram and thermal pressures of the outflow exceed the local magnetic tension.

In Orion~BN/KL, estimates of the magnetic field strength based on the Davis--Chandrasekhar--Fermi method yield values of $\sim$5--10~mG \citep{2017Pattle,2021Cortes}. These measurements indicate that the kinetic energy of the explosive outflow dominates over the magnetic energy, supporting the idea that explosive outflows can overcome magnetic stresses and impose a predominantly radial field configuration. Overall, the observed magnetic field morphologies provided additional evidence that magnetic fields, while dynamically important locally, are strongly shaped by the explosive outflow rather than acting as its primary driver.

\subsection{Gamma-ray emission} \label{sec:gamma}

Recent work has investigated whether explosive outflows in massive star-forming regions can act as particle accelerators capable of producing high-energy gamma rays. Using 16 years of \textit{Fermi}-LAT observations, \citet{Gaches2025} carried out a systematic search for GeV emission toward seven Galactic regions known to host explosive outflows (Orion BN/KL, IRAS 16076, Sh-106, IRAS 12326, DR21, G34.26$+$0.15, and G5.89$-$0.39). They report significant detections associated with DR21, G34.26$+$0.15, and G5.89$-$0.39, while the remaining sources yield only upper limits. Among these, DR21 shows the strongest signal, detected at high significance with a gamma-ray luminosity of order $L_{\gamma}\sim10^{35}$ erg s$^{-1}$. The observed spectra are consistent with models in which cosmic rays are accelerated in shocks driven by the rapidly expanding debris of the explosive outflows as they interact with the surrounding dense molecular gas. By comparing the gamma-ray luminosities with the mechanical energy inferred for the outflows, the authors estimate cosmic-ray acceleration efficiencies of up to $\sim10$--15\%, comparable to those found in other strong astrophysical shocks. These results suggest that explosive outflows produced by dynamical interactions or protostellar mergers in young massive clusters may represent a previously unrecognized class of Galactic gamma-ray sources and could contribute locally to the Galactic cosmic-ray population \citep{Gaches2025}.

\subsection{Hydrogen chloride ($\text{HCl}$)} \label{sec:chlor}

\noindent  \citet{Boehm2026} presented a systematic Atacama Pathfinder Experiment (APEX) survey targeting interstellar hydrogen chloride ($\text{HCl}$) across 28 high-mass star-forming regions, providing an unexpected link to energetic star-formation dynamics. Utilizing the SEPIA660 receiver to observe the $J=1\rightarrow0$ ground-state transitions of $\text{H}^{35}\text{Cl}$ and $\text{H}^{37}\text{Cl}$, the authors apply \texttt{XCLASS} radiative transfer modeling to untangle complex profiles, yielding typical $\text{H}^{35}\text{Cl}$ column densities of $\sim10^{13}\text{ cm}^{-2}$ and localized $[\text{H}^{35}\text{Cl}]/[\text{H}^{37}\text{Cl}]$ ratios between 1.6 and 3.5. They reported a clear correlation of this molecule with the explosive outflows. The detection of $\text{HCl}$ emission within highly energetic and explosive molecular outflows suggests that volatile chlorine chemistry is actively driven by the intense shocks and high temperatures characteristic of sudden, violent dynamical events.

\subsection{Possible origin} \label{sec:origin}

Any viable physical model must simultaneously account for all the reported observational characteristics of explosive outflows. In the following, we summarize the main scenarios that have been proposed in the literature to explain their nature, origin, and driving mechanism. These models can be broadly grouped into three conceptual families:

\begin{itemize}
    \item Models driven by hydrodynamic or magnetohydrodynamic processes acting on pre-existing gas structures,
    \item Models powered by the release of gravitational energy during stellar or protostellar interactions, and
    \item Models invoking catastrophic stellar evolution events.
\end{itemize}

This classification provides a useful framework to assess the physical ingredients required to reproduce the observed morphology, kinematics, and energetics of explosive outflows.

\subsubsection{Free expansion of an H\,II region} \label{sec:instabilities}

More than three decades ago, \citet{Okuda1986} proposed that the Orion outflow resulted from an expanding shell driven by a continuous wind which, through its interaction with the surrounding medium --- and possibly with a slower, previously ejected outflow --- became highly unstable. In this picture, the observed filaments would arise from Rayleigh--Taylor instabilities \citep{Stone1995}. This scenario has since been discarded as the primary origin of explosive outflows, as it requires a long-lived central engine, fails to reproduce the observed ballistic $v\propto r$ kinematics, and predicts a wide range of filament formation timescales that are inconsistent with observations.

Although hydrodynamic instabilities alone cannot explain the global properties of explosive outflows, they are expected to naturally develop during their evolution, once dense clumps propagate supersonically through the ambient medium. In this context, Rayleigh--Taylor and Kelvin--Helmholtz instabilities may play a secondary role in shaping the fine-scale morphology of individual fingers \citep{Betal2015}.

\subsubsection{Magnetic explosion} \label{sec:magnetic}

Another possibility is that the Orion explosion originated from the impulsive release of magnetic energy through large-scale reconnection. In this scenario, magnetic energy stored in a highly magnetized massive core or disk system is suddenly released, inflating a rapidly expanding magnetized bubble that fragments into multiple filamentary structures as a result of MHD instabilities. 

However, current estimates of the available magnetic energy in massive star-forming regions fall well short of the $\sim10^{48}$~erg required to power explosive outflows, unless unrealistically strong magnetic fields ($\gtrsim100$~mG) are invoked over parsec-scale volumes \citep{Pudritz2019}. Moreover, purely magnetic models struggle to reproduce the observed ballistic $v\propto r$ kinematics and do not naturally explain the presence of runaway stars. For these reasons, magnetic explosions are generally considered an auxiliary or contributing mechanism rather than the primary driver of explosive outflows \citep{zap2009,bally2017}.

\subsubsection{Superposition of bipolar outflows} \label{sec:superposition}

The Orion Nebula is a star-forming region densely populated by protostars, which motivated models in which multiple sources inject collimated, jet-like material in different directions to reproduce the apparently isotropic filament morphology \citep{Schultz2001}. However, for a dynamical age of $\sim500$~yr and $\sim100$ bipolar sources, the total injected energy would be of order $10^{46}$~erg, well below the mechanical energy observed in typical explosive outflows (Table~\ref{tab:explosive_outflows_refs_foot}). This scenario would also require an implausibly high source density --- $\sim100$ protostars within a radius of $\sim400$~au, corresponding to a mean density of $\sim3\times10^{9}$~sources~pc$^{-3}$ --- as well as extreme synchronization of jet activity, and it fails to explain the observed runaway stars.

A related model proposed a single powerful jet oriented close to the line of sight that impacts a dense region and produces a quasi-isotropic morphology in projection \citep{Rimmer2012}. However, this model does not account for the observed radial expansion \citep{zap2009}, and a single jet lacks sufficient energy to drive the explosive event.

More recently, interactions between bipolar outflows from close or binary sources have been observed and simulated \citep{Zetal2018,RGetal2025,COHENetal2025}. These interactions tend to disrupt collimation and produce morphologies that are inconsistent with those observed in explosive outflows.

\subsubsection{Failed supernova explosion} \label{sec:supernova}

The explosive morphology and energetics of Orion have also motivated models invoking a failed or partial supernova explosion. In this scenario, an evolved massive star ejects material with a fraction of the canonical supernova energy ($\sim10^{51}$~erg), which is subsequently decelerated by interaction with the surrounding medium, allowing a subset of high-velocity filaments to escape and generate an apparently isotropic outflow.

More recently, \citet{Retal2021} modified this idea by proposing that an evolved massive star hosting hundreds of protoplanets at the onset of core collapse would eject these bodies during the explosion. The resulting high-velocity clumps could form the observed filaments, and if the protoplanets have orbital inclinations randomly distributed within $\pm5^{\circ}$ of the mean plane, the resulting morphology resembles that of observed explosive outflows. In this scenario, the released energy could reach $\sim10^{48}$~erg if the protoplanet masses are on the order of that of Jupiter.

This model is challenged by the absence of nucleosynthetic signatures or a detectable supernova remnant, despite the presence of hot cores and ultracompact H\,II regions, as well as by the need to explain the observed runaway stars without leaving behind a compact remnant.

\subsubsection{Gravitational outflow} \label{sec:gravitational}

While this physical mechanism stands as the most compelling framework to explain the phenomenon of explosive outflows, numerous observational and theoretical components must still be integrated to achieve a comprehensive understanding.

\paragraph{High speed compact object}

This scenario belongs to the family of gravitational interaction models but differs fundamentally from direct stellar collisions or mergers. Instead, it relies on a close gravitational encounter between a high-velocity runaway massive protostar (the ``bullet'') and a self-gravitating system of low-mass clumps.

\begin{figure*}[htbp]
\centering
\includegraphics[width=0.9\linewidth]{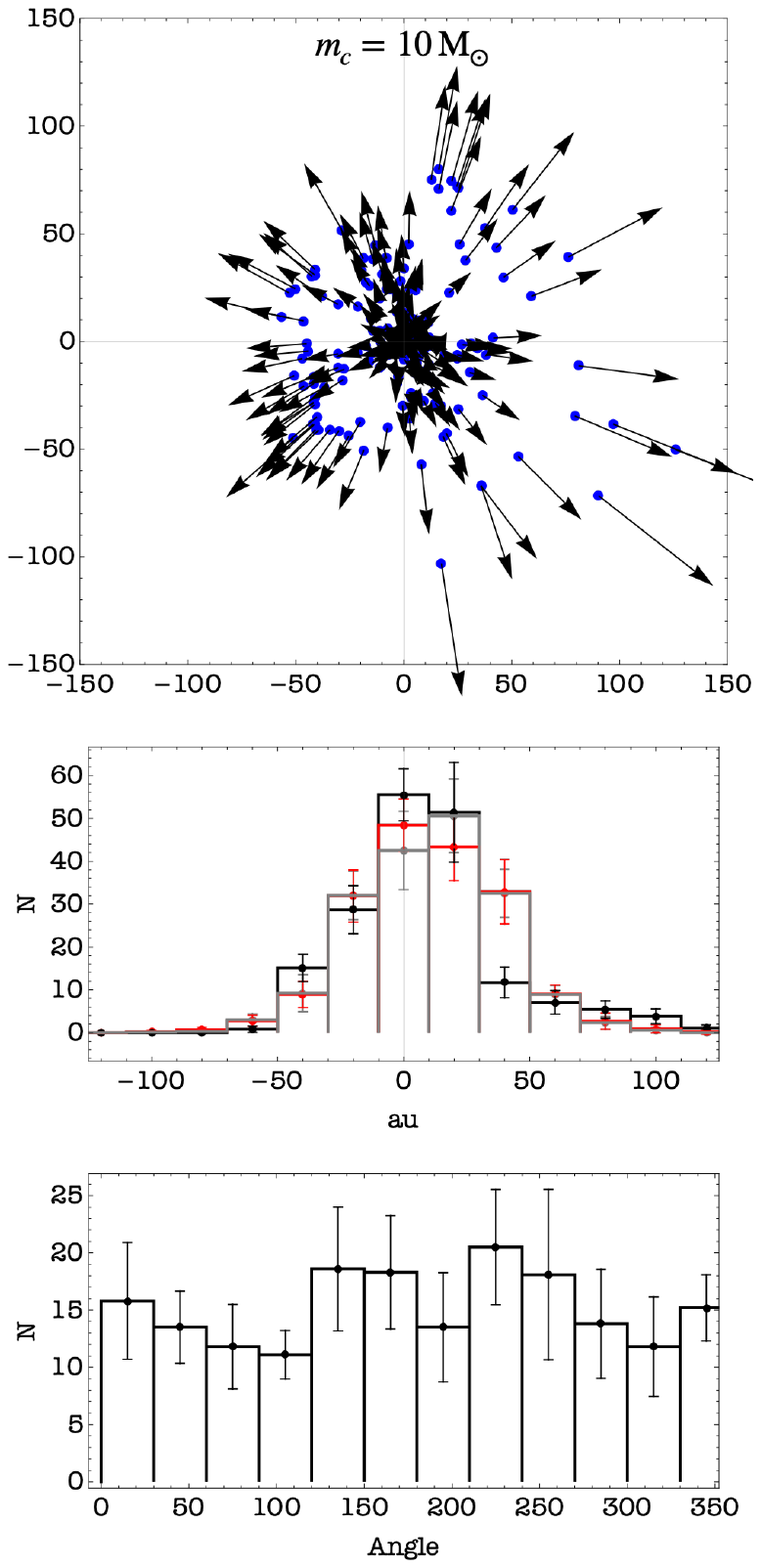}
\caption{
Explosion isotropy from a gravitational outflow in the reference frame of the center of mass of the unbound population. The upper panel shows the spatial distribution of the unbound particles (blue points), with arrows indicating their corresponding velocity vectors. The middle panel presents histograms of particle positions relative to the planes $x = 0$ (black), $y = 0$ (gray), and $z = 0$ (red).
The bottom panel shows the distribution of position angles measured with respect to the $x$-axis. Image reproduced with permission from \citet{ROetal2025}, copyright by the author(s). }
\label{fig:13}
\end{figure*}

The mechanism is analogous to a gravitational slingshot, in which individual clumps are impulsively accelerated during the flyby of the bullet. In the test-particle limit, each clump can acquire velocities up to twice that of the bullet \citep{RO2021}, while for finite substellar clump masses of order $0.05\,M_\odot$, and a total clump mass comparable to that of the bullet, velocities reaching up to three times the bullet speed are possible \citep{ROetal2025}. As a result, an initially compact and isotropic clump distribution is transformed into a radially expanding system obeying $v \propto r$, see Fig.~\ref{fig:13}.

In the case of Orion, this model requires a high-velocity runaway star ($\sim100$\kms) encountering an autogravitating system of Jupiter-mass gas clumps, which are then ejected isotropically at velocities up to $\sim300$\kms from a region only a few astronomical units in size. While such high stellar velocities are rare, observational evidence indicates that they do occur \citep{Gvaramadze2009}.

This framework has been explored quantitatively by \citet{RO2019a,RO2019b}, who modeled each filament as a mass-losing plasmon ejected during a single impulsive event. Their results successfully reproduce the observed kinematics, constrain the clump masses and initial velocities, and predict characteristic lifetimes of order $10^3$~yr for explosive outflows such as Orion. At present, gravitationally induced explosive outflows provide one of the most self-consistent explanations for the observed isotropy, energetics, ballistic expansion, and the simultaneous presence of runaway stars, although open questions remain regarding the origin and mass spectrum of the clumps involved. Moreover, there is no evidence of the high-velocity runaway star required by the model.

\paragraph{Protostellar mergers} \label{sec:mergers}

This scenario falls within the broader class of gravitational interaction models. 

In dense young clusters, dynamical interactions can produce tight binaries whose eventual merger provides a viable pathway for massive star formation. Binary-mediated mergers can occur at stellar densities typical of young massive clusters, naturally overcoming the radiation pressure problem, while releasing large amounts of gravitational energy ($\sim10^{47}$--$10^{48}$~erg) capable of driving powerful outflows and disrupting the surrounding medium \citep{2005bally,2005bon}. This scenario further predicts that the most massive stars are likely to be single merger products, consistent with observations of massive stars in dense cluster environments.

It is highly probable that the material was ejected along ballistic trajectories from the disks or envelopes of the runaway stars during the dynamical encounter. This debris is observed today in the form of high-velocity bullets ($\sim \text{a few hundred}~\mathrm{km~s^{-1}}$) that structurally constitute the explosive outflow.
  
\newpage
\subsubsection{Summary}

In conclusion, all these proposed scenarios for the explosive molecular outflows may represent the dynamical aftermath of extreme interactions in dense young stellar clusters, potentially involving close encounters or mergers among massive protostars. In this picture, the rapid release of gravitational energy ($\sim10^{47}$–$10^{48}$ erg) could disrupt circumstellar material and drive fast, wide-angle ejecta with roughly Hubble--Lemaître-like expansions, although other transient mechanisms cannot be ruled out. The short dynamical ages and high energies of these outflows suggest a brief, non-steady origin that differs from classical jet-driven feedback and trace a distinct, previously unidentified phase in the unfolding of the massive star formation process.

\section{Main conclusions}

The formation of massive stars ($\gtrsim8\,M_\odot$) remains a central problem in star formation theory because it proceeds rapidly, deeply embedded within dense molecular clouds, and under conditions where stellar feedback is expected to strongly oppose continued mass accretion. Massive stars are observed to form preferentially in clustered environments, often at the centers of young massive clusters, where gas densities, stellar densities, and dynamical interactions are all enhanced. Proposed formation scenarios include disk-mediated accretion onto massive protostars, competitive accretion within cluster potentials, and the merging of intermediate- and low-mass protostars during dynamical interactions. Discriminating among these mechanisms requires identifying observational signatures that can directly trace the physical processes operating during the earliest, most deeply embedded phases of massive star formation.

\begin{enumerate}

\item Explosive molecular outflows, as characterized in Sect.~\ref{sec:common}, trace remarkable, energetic, and short-lived events with dynamical ages of $\sim10^{2}$--$10^{3}$~yr, extreme velocities reaching $\gtrsim100$\kms, quasi-isotropic configurations with molecular streamers following Hubble--Lemaître-like expansions, and kinetic energies of $E_\mathrm{kin}\sim10^{46}$--$10^{48}$~erg. These properties indicate a transient, impulsive origin that is fundamentally distinct from the steady, jet-driven feedback associated with classical bipolar molecular outflows.

\item  As discussed in Sect.~\ref{sec:bolo}, the strong association of explosive outflows with dense and luminous cluster environments supports a close connection with dynamical interactions such as close encounters, binary hardening, and protostellar mergers. These processes are expected to operate efficiently at stellar densities of $\sim10^{6}$--$10^{8}$~pc$^{-3}$ during brief phases of cluster assembly, conditions that can be reached in the cores of young massive clusters and during the collapse of dense, fragmenting molecular cores.

\item  Explosive outflows are distributed across a wide range of Galactocentric radii, indicating that the conditions required for their formation are not limited to extreme environments near the Galactic center. Independent rate estimates based on spatial scaling and steady-state arguments consistently yield a Galactic occurrence rate of $\sim7 \times10^{-3}$~yr$^{-1}$, corresponding to roughly one event every $\sim140$~yr. Although these estimates are affected by observational biases, they suggest that explosive outflows are not rare events that are closely linked to the formation of massive stars.

\item  Explosive outflows exhibit mechanical luminosities that are dramatically higher than those of classical protostellar outflows, reaching $L_\mathrm{mech}\sim10^{36}$--$10^{39}$~erg~s$^{-1}$ over short dynamical timescales of $\sim10^{3}$--$10^{4}$~yr. For a given bolometric luminosity, their mechanical output exceeds that of steady, accretion-driven outflows by one to three orders of magnitude, placing them well above the empirical $L_\mathrm{mech}$--$L_\mathrm{bol}$ relation followed by classical systems. This clear energetic excess strongly disfavors continuous jet or wind scenarios and instead supports an impulsive origin, such as stellar dynamical interactions or protostellar mergers, capable of releasing large amounts of energy over short timescales.

\item The diversity of models reviewed in Sect.~\ref{sec:origin} highlights that explosive outflows likely arise from extreme and short-lived events in dense star-forming environments, and that reproducing their full set of observed properties places strong constraints on the underlying physical mechanism. While several scenarios can account for individual aspects of the phenomenon, only those involving impulsive gravitational interactions are able to simultaneously explain the isotropic filamentary morphology, ballistic $v\propto r$ kinematics, large kinetic energies, and the presence of runaway stars. 

\item Millimeter polarization observations toward Orion~BN/KL, G5.89$-$0.39, and G34.26$+$0.15 reveal predominantly radial magnetic field morphologies associated with explosive outflows, when interpreted under the assumption of dust grains aligned with the magnetic field. While alternative alignment mechanisms may contribute locally, the large-scale field structure appears closely linked to the outflow geometry, extending down to radii of $\sim$1{,}500~au from the explosion centers. The observed transition from irregular or azimuthal configurations in dense structures to radial fields at larger radii is consistent with magnetic flux-freezing, in which field lines are compressed or dragged outward by the expanding flow. Estimates of the magnetic field strength ($\sim$5--10~mG in Orion~BN/KL) indicate that the kinetic energy of the explosive outflows dominates over the magnetic energy, suggesting that magnetic fields are shaped by the outflow rather than acting as its primary driver.

\item The implications of explosive outflows for cluster evolution and stellar mass assembly, suggest that such events may preferentially enhance the growth of the most massive cluster members while dynamically ejecting or suppressing lower-mass stars. This combination of processes may bias the stellar initial mass function toward a top-heavy form in the densest cluster cores, linking explosive outflows to the origin of unusually massive stars and compact stellar systems.

\item Finally, the observational tests and future prospects including runaway or walkaway stars with proper motions tracing back to the explosion center, a reduced close-binary fraction or an excess of apparently single massive stars, truncated or disrupted circumstellar disks near the outflow origin, strong mass segregation on spatial scales comparable to the outflow launch site, and a statistically top-heavy IMF in the densest regions. High-angular-resolution and high-sensitivity observations with ALMA and the ngVLA will be essential to evaluate these predictions and to establish the role of explosive outflows in regulating cluster formation and the high-mass end of the stellar initial mass function.

\end{enumerate}

\backmatter

%\bmhead{Supplementary information}

%If your article has accompanying supplementary file/s please state so here. 

%Authors reporting data from electrophoretic gels and blots should supply the full unprocessed scans for key as part of their Supplementary information. This may be requested by the editorial team/s if it is missing.

%Please refer to Journal-level guidance for any specific requirements.

\bmhead{Acknowledgements}

I am deeply grateful to Dr. Manuel López Fernández for his invaluable help and insightful contributions to the Magnetic Fields in Explosive Outflows section, and to Dr. Pedro Rubén Rivera for many stimulating discussions and his key contributions to the Possible Origin of Explosive Outflows section. I also thank Dr. Luis Felipe Rodríguez and Dr. Paul Ho for their thoughtful revisions and inputs. Finally, I also thank the anonymous referees for their helpful comments and constructive review, which significantly improved the clarity of the manuscript.

\bmhead{Funding}
L.A.Z. acknowledges financial support from CONACyT grant 280775 and UNAM–PAPIIT grants IN110618 and IN112323 (Mexico).

\section*{Declarations}

\bmhead{Conflict of interest} The author declares no conflict of interest.

%\begin{appendices}

%\section{Section title of first appendix}\label{secA1}

%An appendix contains supplementary information that is not an essential part of the text itself but which may be helpful in providing a more comprehensive understanding of the research problem or it is information that is too cumbersome to be included in the body of the paper.




%\end{appendices}


\phantomsection
\addcontentsline{toc}{section}{References}
\bibliography{sn-bibliography.bib}% common bib file

\end{document}